\documentclass[twocolumn]{aastex62}

\usepackage{multirow,arydshln,hhline}

\submitjournal{ApJ}

\shorttitle{OGLE Cepheids in the Magellanic Bridge}
\shortauthors{Jacyszyn-Dobrzeniecka et al.}

\begin{document}

\title{OGLE-ING THE MAGELLANIC SYSTEM: CEPHEIDS IN THE BRIDGE\footnote{Draft version prepared on April, 16th, 2019}}

\correspondingauthor{Anna M. Jacyszyn-Dobrzeniecka}
\email{jacyszyn@uni-heidelberg.de}

\author[0000-0002-5649-536X]{Anna M. Jacyszyn-Dobrzeniecka}
\affiliation{Astronomical Observatory, University of Warsaw, Al. Ujazdowskie 4, 00-478 Warszawa, Poland}
\affiliation{Astronomisches Rechen-Institut, Zentrum f\"ur Astronomie der Universit\"at Heidelberg, M\"onchhofstr. 12-14, D-69120 Heidelberg, Germany}

\author{Igor Soszy\'nski}
\affiliation{Astronomical Observatory, University of Warsaw, Al. Ujazdowskie 4, 00-478 Warszawa, Poland}

\author{Andrzej Udalski}
\affiliation{Astronomical Observatory, University of Warsaw, Al. Ujazdowskie 4, 00-478 Warszawa, Poland}

\author{Micha\l{} K. Szyma\'nski}
\affiliation{Astronomical Observatory, University of Warsaw, Al. Ujazdowskie 4, 00-478 Warszawa, Poland}

\author{Dorota M. Skowron}
\affiliation{Astronomical Observatory, University of Warsaw, Al. Ujazdowskie 4, 00-478 Warszawa, Poland}

\author{Jan Skowron}
\affiliation{Astronomical Observatory, University of Warsaw, Al. Ujazdowskie 4, 00-478 Warszawa, Poland}

\author{Przemek Mr\'oz}
\affiliation{Astronomical Observatory, University of Warsaw, Al. Ujazdowskie 4, 00-478 Warszawa, Poland}

\author{Katarzyna Kruszy\'nska}
\affiliation{Astronomical Observatory, University of Warsaw, Al. Ujazdowskie 4, 00-478 Warszawa, Poland}

\author{Patryk Iwanek}
\affiliation{Astronomical Observatory, University of Warsaw, Al. Ujazdowskie 4, 00-478 Warszawa, Poland}

\author{Pawe\l{} Pietrukowicz}
\affiliation{Astronomical Observatory, University of Warsaw, Al. Ujazdowskie 4, 00-478 Warszawa, Poland}

\author[0000-0002-9245-6368]{Rados\l{}aw Poleski}
\affiliation{Department of Astronomy, Ohio State University, 140 West 18th Avenue, Columbus, OH 43210, USA}

\author{Szymon Koz\l{}owski}
\affiliation{Astronomical Observatory, University of Warsaw, Al. Ujazdowskie 4, 00-478 Warszawa, Poland}

\author{Krzysztof Ulaczyk}
\affiliation{Department of Physics, University of Warwick, Coventry CV4 7AL, UK}

\author{Krzysztof Rybicki}
\affiliation{Astronomical Observatory, University of Warsaw, Al. Ujazdowskie 4, 00-478 Warszawa, Poland}

\author{Marcin Wrona}
\affiliation{Astronomical Observatory, University of Warsaw, Al. Ujazdowskie 4, 00-478 Warszawa, Poland}

\begin{abstract}

We present a detailed analysis of Magellanic Bridge Cepheid sample constructed using the OGLE Collection of Variable Stars. Our updated Bridge sample contains 10 classical and 13 anomalous Cepheids. We calculate their individual distances using optical period--Wesenheit relations and construct three-dimensional maps. Classical Cepheids on-sky locations match very well neutral hydrogen and young stars distributions, thus they add to the overall Bridge young population. In three dimensions, eight out of ten classical Cepheids form a bridge-like connection between the Magellanic Clouds. The other two are located slightly farther and may constitute the Counter Bridge. We estimate ages of our Cepheids to be less than 300 Myr for five up to eight out of ten, depending on whether the rotation is included. This is in agreement with a scenario where these stars were formed \textit{in-situ} after the last encounter of the Magellanic Clouds. Cepheids' proper motions reveal that they are moving away from both Large and Small Magellanic Cloud. Anomalous Cepheids are more spread than classical Cepheids in both two and three dimensions. Even though, they form a rather smooth connection between the Clouds. However, this connection does not seem to be bridge-like, as there are many outliers around both Magellanic Clouds.

\end{abstract}



\section{Introduction} \label{sec:intro}

The Magellanic Bridge (MBR), which undoubtedly is a direct evidence of the Magellanic Clouds' interactions, has been a subject of interest of many research projects. Though observations of the Bridge area started with Shapley's first discovery of young stars located in the SMC Wing \citep{Shapley1940}, the Bridge as a structure was discovered as a hydrogen feature \citep{Hindman1963}. Numerical models predict that the connection between the Large and Small Magellanic Cloud (LMC and SMC, respectively) was formed after their last encounter, about $200-300$ Myr ago (e.g. \citealt{Gardiner1994,Gardiner1996,Ruzicka2010,Diaz2012,Besla2012}) or, as recent study shows, slightly later -- about 150 Myr ago \citep{Zivick2019}.

Different studies of the gaseous counterpart of the MBR showed that it is a rather complicated, multi-phase structure (\citealt{DOnghia2016} and references therein). The neutral hydrogen (\textsc{H i}) kinematics reveal that the Bridge is connected with the western parts of the LMC disk \citep{Indu2015}, and moreover, is also being sheared. Other studies showed that the Bridge also contains warm ionized gas \citep{Barger2013}. Moreover, \citet{WagnerKaiser2017} found evidence of dust in the MBR, concluding that it has probably been pulled out of either or both Clouds during their interactions.

Here we present a detailed analysis of classical and anomalous Cepheids in the Bridge area. Different stellar components of the Bridge have been discovered up to date. This is in agreement with numerical models predictions (e.g. \citealt{Diaz2012,Besla2012,Guglielmo2014}). Many studies were devoted to searching for young stars between the Magellanic Clouds and found an evidence of their presence therein \citep{Shapley1940,Irwin1985,Demers1998,Harris2007,Noel2013,Noel2015,Skowron2014,Belokurov2017,Mackey2017,Zivick2019}. \citet{Skowron2014} showed using the Optical Gravitational Lensing Experiment (OGLE) data that young stars form a continuous bridge-like connection and their distribution is clumped. This was confirmed by \citet{Belokurov2017} who tested young main sequence stars from {\it Gaia} and GALEX, as well as \citet{Mackey2017} who used Dark Energy Camera data. Young ages of some of these stars strongly suggest an \textit{in-situ} formation. \citet{Zivick2019} found a correlation between the young population and \textsc{H i}. Moreover, studies of stellar proper motions for both young and old population \citep{Oey2018,Zivick2019} show that the Bridge is moving away from the SMC towards the LMC.

The clumped pattern of stellar associations distribution between the Magellanic Clouds may suggest an ongoing process of forming a tidal dwarf galaxy \citep{Bica1995,Bica2015,Ploeckinger2014,Ploeckinger2015,Ploeckinger2018}. Recently, a dwarf galaxy was found located in the on-sky Bridge area, though it is located halfway between the Sun and the Magellanic System \citep{Koposov2018}.

Classical pulsators were also studied in the MBR. \citet{Soszynski2015b}, as part of the OGLE Collection of Variable Stars (OCVS), published a list of classical Cepheids (CCs) including new discoveries located in the MBR. Jacyszyn-Dobrzeniecka et al. (2016; hereafter \citealt{PaperI}) studied their three-dimensional distribution and classified nine as MBR members. Five of these objects seem to form a bridge-like connection between the Magellanic Clouds, while the others are more spread in three-dimensions. Ages of these CCs suggest that they were formed \textit{in-situ}, as almost all are under 300 Myr.

The evidence was found for intermediate-age and old stars between the Magellanic Clouds \citep{Bagheri2013,Noel2013,Noel2015,Skowron2014,Carrera2017}. Classical pulsators belonging to the latter group, the RR Lyrae stars, are also present in the MBR and their distribution was thoroughly tested (Jacyszyn-Dobrzeniecka et al. 2017, hereafter \citealt{PaperII,WagnerKaiser2017,Belokurov2017}). Also Mira candidates were searched for in the MBR \citep{Deason2017}. Another paper in the series of using OCVS to analyze the three-dimensional structure of the Magellanic System (Jacyszyn-Dobrzeniecka et al. 2019, hereafter \citealt{PaperIV}), following closely this paper, summarizes and updates the current knowledge of RR Lyrae stars distribution in the Bridge. For more information on the old stellar counterpart of the MBR see Introduction in \citealt{PaperIV}.

In this work we present an analysis of Cepheids in the Magellanic Bridge using the updated, corrected and extended OGLE data. We studied three-dimensional distributions of classical Cepheids (CCs), anomalous Cepheids (ACs), as well as type II Cepheids (T2Cs), though we did not classify any of the latter as MBR members. For CCs and ACs we also present a detailed analysis of many parameters and a comparison of different methods used. In this paper we also compare our sample to {\it Gaia} Data Release 2 (DR2) Cepheids and for the first time present their distribution in the Bridge.

We organized the paper as follows. In Section~2 we present the OCVS as well as the latest changes and updates applied to the Collection. Section~3 presents methods of calculating individual distances and coordinates transformation. A detailed analysis of CCs and ACs distributions is included in Sections~4 and 5, respectively. In Section~6 we discuss the influence of the recent reclassification of four Cepheids on their parameters. For the first time we present {\it Gaia} Data Release 2 (DR2) Cepheids in the Bridge and compare them to the OCVS Cepheids in Section 7. We summarize and conclude the paper in Section~8.


\section{Observational Data}

\subsection{OGLE Collection of Variable Stars} \label{sec:ocvs}

\begin{figure*}[htb]
	\includegraphics[width=\textwidth]{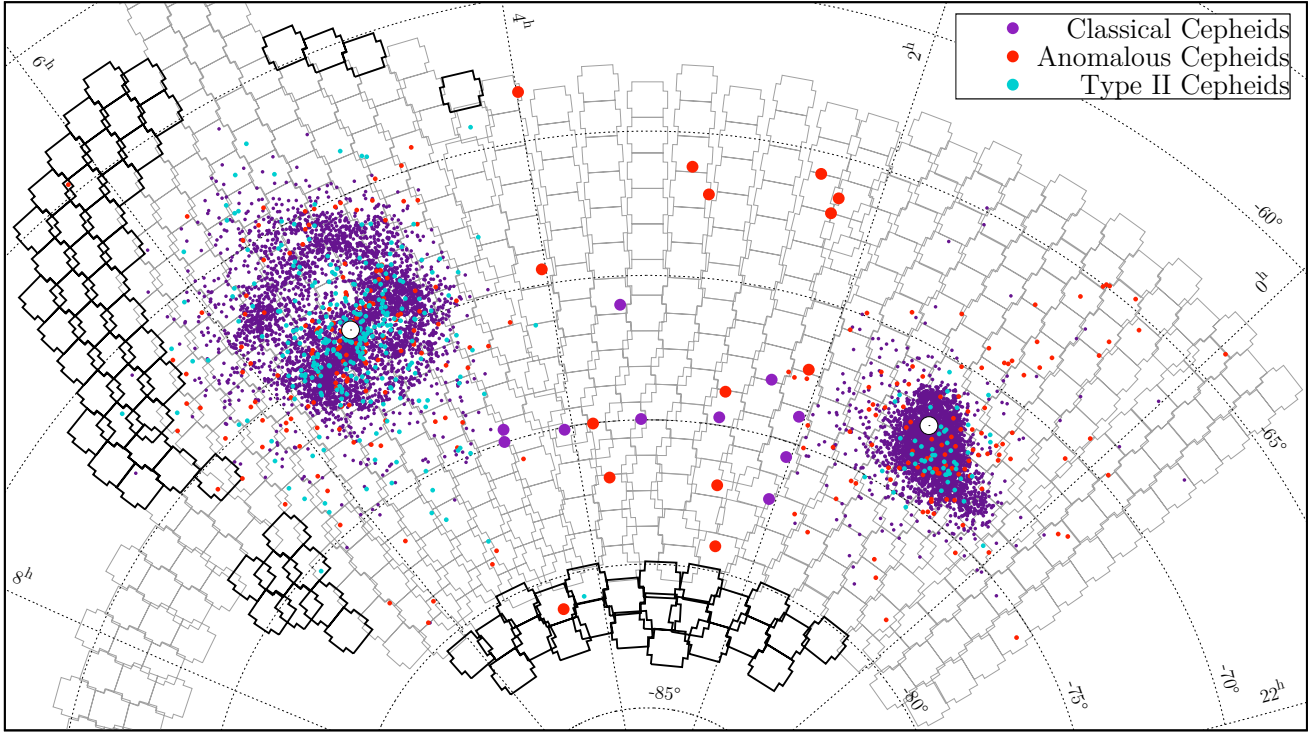}
	\caption{The on-sky locations of Cepheids in the Magellanic System. The selected Bridge sample is featured with larger dots. Black contours show the newest addition to the OGLE-IV fields while grey show main OGLE-IV fields in the Magellanic System that were already observed before July 2017. White circles mark the LMC \citep{vanderMarel2014} and SMC \citep{Stanimirovic2004} centers.}
	\label{fig:cep-all}
\end{figure*}

In this study we use data from the fourth phase of the OGLE project \citep{Udalski2015}. In particular, we use Cepheids from the OCVS \citep{Soszynski2015b,Soszynski2017} as well as new unreleased data in the Magellanic System. Most of the updates come from the newly added OGLE fields that are marked with black contours in Fig.~\ref{fig:cep-all}. The extension of the MBR fields is located in the southern parts of the Magellanic System. There are also many fields added in the northern and eastern parts of the LMC. The numbers of Cepheids that were lately added to the OCVS is presented in Tab.~\ref{tab:cep-add}. These numbers are very low and this is what we expect since OGLE collection of Cepheids was already nearly complete \citep{Soszynski2015b,Soszynski2017}.

\setlength\dashlinedash{1pt}
\setlength\dashlinegap{2pt}

\begin{table}[htb]
\caption{Latest additions to the OGLE Collection of Cepheids in the Magellanic System}
\label{tab:cep-add}
\centering
\begin{tabular}{lDDDD}

Source & \multicolumn2c{CCs F} & \multicolumn2c{ACs F} & \multicolumn2c{ACs 1O} & \multicolumn2c{All} \\
\cline{1-9}
\decimals
New MBR fields$^{(a)}$   &  - & - & 1 & 1 \\
New LMC fields$^{(a)}$   &  2 & 1 & - & 3 \\
{\it Gaia} DR2$^{(b)}$         &  2 & 1 & - & 3 \\
\cdashline{1-9}
All                      &  4 & 2 & 1 & 7 \\
\cline{1-9}
\multicolumn{9}{p{.45\textwidth}}{F stands for the fundamental mode, whereas 1O for the first-overtone pulsators. $(a)$ For a current OGLE-IV footprint with these newly added fields see Fig.~\ref{fig:cep-all}. $(b)$ We searched for Cepheids that are present in {\it Gaia} DR2 but not in OCVS. After careful studies of their OGLE light curves we classified a few additional stars in the Magellanic System.}

\end{tabular}
\end{table}

Tab.~\ref{tab:cep-add} also includes three newly confirmed Cepheids based on {\it Gaia} DR2 \citep{Gaia2018}. We used their publicly available list of Cepheids \citep{Holl2018,Clementini2019} to cross-match with the OCVS and searched for objects that were not present in the latter. After a careful inspection of their OGLE light curves, we identified three Cepheids that we included in a sample analyzed in this paper.

Moreover, we reclassified types and modes of pulsation for four Cepheids from the MBR area that were presented in \citealt{PaperI}. This is due to their light curve Fourier decomposition parameters suggesting different classification \citep{Soszynski2015a}. One Cepheid was moved from first-overtone to fundamental mode CC. Three CCs were reclassified as ACs.

The public version of the OCVS is soon to be updated with the changes described in this section (Soszy\'nski et al. in prep.).

For one CC in our sample, namely OGLE-SMC-CEP-4986, the $V$-band magnitude was not available in the OGLE database. Thus, we used the ASAS-SN Sky Patrol light curve \citep{Schappee2014,Kochanek2017} to calculate its mean magnitude in the $V$-band. To make sure it is properly calibrated, we selected 10 reference stars located in the same detector (OGLE operates a 32-chip mosaic camera) as the Cepheid. These objects were non-variables and had the closest magnitude and color to the OGLE-SMC-CEP-4986, as well as good quality magnitude measurement in the OGLE database (many epochs). For the reference stars we compared magnitudes in the OGLE and ASAS-SN Sky Patrol and calculated a correction, which was at the order of 0.08 mag.


\section{Data Analysis}

\subsection{Period--Luminosity Relations and Individual Distances} \label{sec:plr-dist}

To calculate individual distances of Cepheids we used the entire Magellanic System samples and applied the same technique as in \citealt{PaperI} (see Section~3.1 therein for more details). We did this separately for CCs and ACs. Using  Wesenheit magnitudes \citep{Madore1982}, we fitted period--luminosity (PL) relations (Leavitt Law) to the LMC sample (see Eqs.~1 and 2 in \citealt{PaperI}). Together with the least-square method we applied $3\sigma$ clipping to the data. We note that, however, this approach may not be the most appropriate for studying distances \citep{Deb2018}, as \citet{Nikolaev2004} showed that the error distribution is not normal for Wesenheit index at a given period. On the other hand, many studies proved this technique to be very robust in case of the Magellanic System (i.e. \citealt{Haschke2012a,Haschke2012b, Moretti2014,PaperI,Inno2016,Ripepi2017}).

\begin{figure*}[htb]
	\begin{center}
	\includegraphics[width=\textwidth]{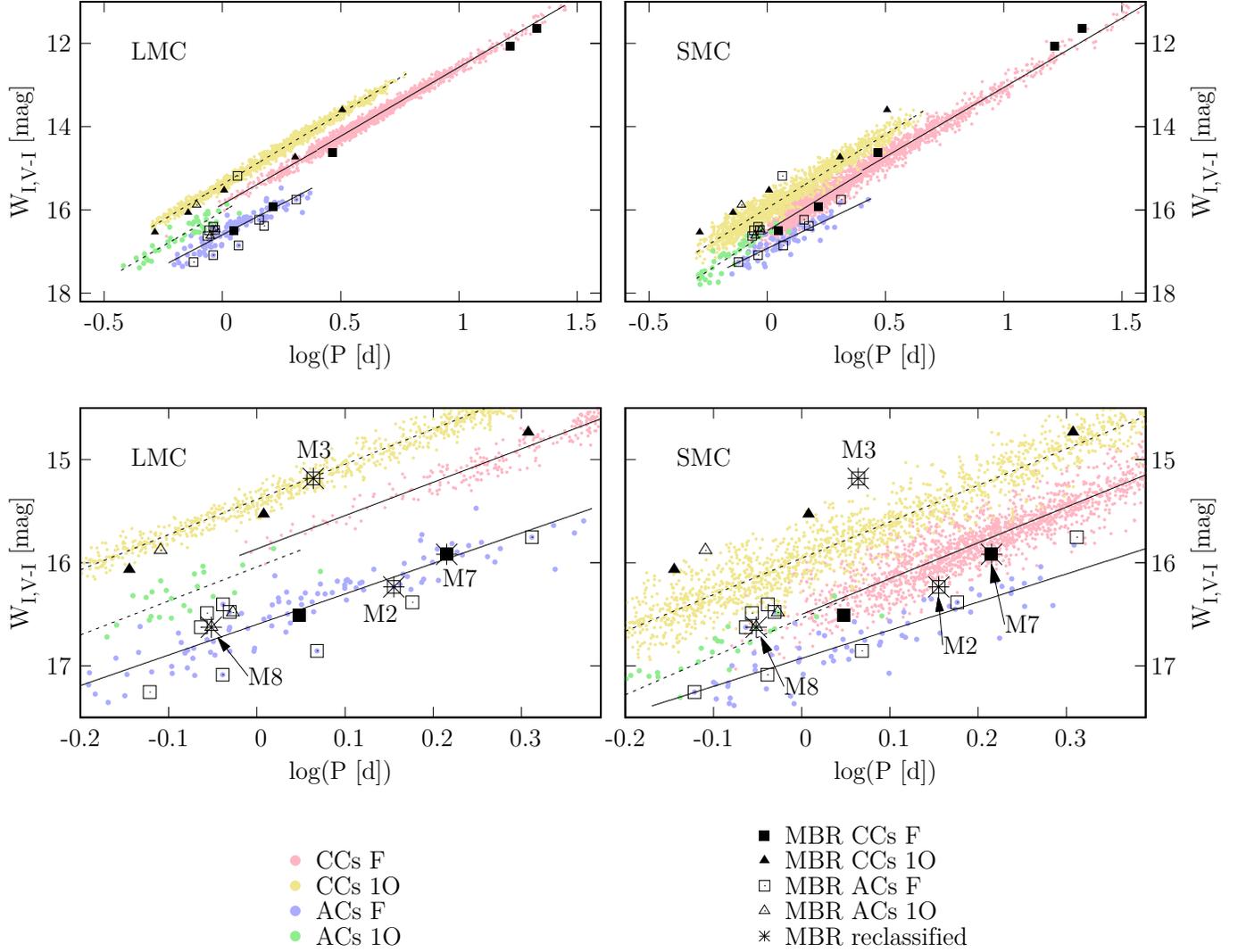}
	\end{center}
	\caption{Period--luminosity relations for classical and anomalous Cepheids in the LMC (left column) and SMC (right column). CCs are marked with smaller dots than ACs. The entire Bridge sample is overplotted on the presented PL relations in every panel with each type marked separately. Additionally, bottom panels highlight four Cepheids that were reclassified and are marked with a star and their local ID. M7 was reclassified from first-overtone CC to fundamental mode CC, M2 and M3 -- from fundamental mode CCs to fundamental mode ACs, and M8 -- from first-overtone CC to first-overtone AC. Plots do not show $3\sigma$ outliers as these were removed from the final sample. The fit for fundamental mode ACs in the SMC has significantly different slope than all of the other relations. Note that, however, we do not use the SMC ACs PL relations and these are only plotted here for comparison.}
	\label{fig:cep-plr}
\end{figure*}

For fundamental mode CCs we included a break in the PL relation at $\log P=0.4$. For first-overtone CCs we excluded objects with $\log P<-0.3$ (see Section~3.1 in \citealt{PaperI} and \citealt{Soszynski2008}). Fig.~\ref{fig:cep-plr} shows separate PL relations for the final LMC and SMC CCs and ACs samples with Bridge Cepheids overplotted on each panel using larger marks. Each type and mode is plotted using a different point type. Additionally, the bottom row highlights the four reclassified Cepheids and shows their local IDs (labels consisting of an ''M'' with a number that we started using in \citealt{PaperI}). The parameters of our fits are consistent with those from \citealt{PaperI} and are shown in Tab.~\ref{tab:ccs-plr}. Number of stars included in the fits is slightly smaller than in \citealt{PaperI} because this time we did not complement our final set with OGLE-III observations.

\begin{table*}[htb]
\caption{PL relations for CCs in the Magellanic System in the Wesenheit magnitude}
\label{tab:ccs-plr}
\centering
\begin{tabular}{ccccccccc}
\multicolumn{9}{c}{$W_{I,V-I}=a\log P+b$} \\
\cline{1-9}
Galaxy & P. mode & $\log P$ & $a$ & $b$ [mag] & $\sigma$ [mag] & $\chi^2$/dof & $N_{\rm inc}$ & $N_{\rm rej}$ \\
\cline{1-9}

\decimals
\multirow{4}{*}{LMC} & \multirow{3}{*}{F} & $\leq 0.4$ & $-3.234\pm0.033$ & $15.866\pm0.010$ & 0.104 & 3.029 &  273 &  6 \\
{                  } & {                } & $>0.4$     & $-3.315\pm0.008$ & $15.888\pm0.005$ & 0.076 & 1.613 & 2042 & 85 \\
{                  } & {                } & all        & $-3.311\pm0.006$ & $15.885\pm0.004$ & 0.079 & 1.714 & 2308 & 98 \\ \cdashline{2-9}
{                  } & 1O                 & all        & $-3.411\pm0.007$ & $15.387\pm0.003$ & 0.077 & 1.634 & 1772 & 85 \\
\cline{1-9}
\multirow{4}{*}{SMC} & \multirow{3}{*}{F} & $\leq 0.4$ & $-3.470\pm0.015$ & $16.501\pm0.004$ & 0.162 & 7.362 & 1698 & 38 \\
{                  } & {                } & $>0.4$     & $-3.330\pm0.008$ & $16.389\pm0.006$ & 0.149 & 6.170 &  935 & 28 \\
{                  } & {                } & all        & $-3.453\pm0.005$ & $16.489\pm0.002$ & 0.159 & 7.106 & 2636 & 63 \\ \cdashline{2-9}
{                  } & 1O                 & all        & $-3.535\pm0.007$ & $15.957\pm0.002$ & 0.171 & 8.198 & 1879 & 30 \\
\cline{1-9}
\multicolumn{9}{p{.73\textwidth}}{$N_{\rm inc}$ is the number of objects included in the fit, while $N_{\rm rej}$ is the number of objects rejected during $3\sigma$-clipping procedure.}
\end{tabular}
\end{table*}

We then followed our previous technique as described in details in Section~3.2 in \citealt{PaperI}. We assumed that the fitted PL relation corresponds to the mean LMC distance and the individual distances were calculated in respect to the best fit (see Eqs.~3, 4, 5 in \citealt{PaperI}). As a reference distance we have used the most accurate result up-to-date obtained by \citet{Pietrzynski2019}. The resulting three-dimensional distribution of CCs is discussed in the next section.

\subsection{Coordinate Transformations}

In this study we again use Hammer equal-area sky projection as we did in \citealt{PaperI} and \citealt{PaperII}. The projection is rotated so that the {\it z} axis is pointing toward $\alpha_{\rm cen}=3^{\rm h}20^{\rm m}, \delta_{\rm cen}=-72^{\circ}$. This time we have introduced one small correction to Eqs.~7--11 from \citealt{PaperI} that leads to a coordinate system with an $x$ axis that is symmetrical with respect to $\alpha_{\rm cen}$. We have also added a coefficient of $-\frac{\pi}{2}$ when normalizing $l$ that was missing in our original equations.
\begin{eqnarray}
	\alpha_{\rm b} &=& \alpha+\left(\frac{\pi}{2}-\alpha_{\rm cen}\right) \\
	l &=& \arctan\left(\frac{\sin (\alpha_{\rm b})\cos (\delta_{\rm cen})+\tan(\delta)\sin (\delta_{\rm cen})}{\cos (\alpha_{\rm b})}\right) \\
	\beta &=& \arcsin(\sin (\delta)\cos (\delta_{\rm cen})-\cos (\delta)\sin (\delta_{\rm cen})\sin (\alpha_{\rm b}))
\end{eqnarray}
$l$ and $\beta$ are auxiliary variables. We normalize the coordinates so that $l-\frac{\pi}{2} \in (-\pi,\pi)$ and $\beta \in \left(-\frac{\pi}{2},\frac{\pi}{2}\right)$.
\begin{eqnarray}
	x_{\rm Hammer} &=& -\frac{2\sqrt{2}\cos (\beta)\sin \left(l/2\right)}{\sqrt{1+\cos (\beta)\cos \left(l/2\right)}} \\
	y_{\rm Hammer} &=& \frac{\sqrt{2}\sin (\beta)}{\sqrt{1+\cos (\beta)\cos \left(l/2\right)}}.
\end{eqnarray}


\section{Classical Cepheids}

\subsection{Updated Bridge Sample}

In this Section we present a detailed analysis of the updated sample of classical Cepheids (CCs) in the Magellanic System in the context of the Magellanic Bridge. The sample of Bridge CCs was first presented by \citet{Soszynski2015b} and included five objects. Later, in \citealt{PaperI} we have enlarged that sample to nine and discussed their three-dimensional locations in details (see Section~6 therein). We labeled the objects M1--M9 (see Tab.~10 in \citealt{PaperI}). Since then, \citet{Soszynski2017} have already added one classical Cepheid to the OGLE Bridge sample making it the tenth one (M10).

In Section~\ref{sec:ocvs} we described the updates and corrections that were applied to the OGLE Collection of Variable Stars very recently. We reclassified M7 from first-overtone CC to fundamental mode CC. We also moved three objects from CCs sample to ACs, namely M2, M3 and M8. The corrections applied influenced Cepheids' distances decreasing them by even up to $\sim 20$ kpc. Thus, the three-dimensional distribution of the Bridge sample has significantly changed as compared to \citealt{PaperI}.

We have constructed our final Bridge Cepheid sample based on the on-sky as well as three-dimensional locations of Cepheids in relation to the LMC and SMC entire samples. We decided to add to the Bridge sample two objects located close to the LMC (M12 and M13). These CCs were already included in the first OGLE-IV Collection of classical Cepheids by \citet{Soszynski2015b}, though we did not incorporate these in \citealt{PaperI} sample. However, having now two CCs located close to the SMC (M9 and M11, M11 was added by \citealt{Soszynski2017} and was not present in \citealt{PaperI} sample) we think it is worth comparing the Bridge sample to the LMC outliers as well. All of these four Cepheids, both located on the SMC side (M9, M11) and on the LMC side (M12, M13) are connecting the Clouds' samples to the genuine MBR sample.

Due to these updates and corrections our final Bridge CCs sample consists of 10 objects. The list of CCs and their basic parameters is included in Tab.~\ref{tab:ccs-gen}, which provides the object's OCVS ID, local ID used in \citealt{PaperI} and this work, pulsation period $P$, mean magnitudes from both OGLE passbands ($I$ and $V$), Right Ascension and Declination (epoch J2000.0), distance $d$ (details on the method used -- see Section~\ref{sec:plr-dist}), and age estimated using the period--age relation from \citet{Bono2005}. The list comprises of five fundamental mode, four first-overtone pulsators and one double-mode Cepheid (pulsating simultaneously in the first and second overtone) for which we used its first-overtone period in this analysis.

Our Bridge Cepheid sample consists also of ACs that we discuss in Sec.~\ref{sec:acs}. We also note that we did not classify any of the recently published T2Cs in the Magellanic System \citep{Soszynski2018} as a Bridge candidate, as these stars do not seem to form any bridge-like connection and none is located in the direct area of interest.

It is noteworthy, however, that \citet{Iwanek2018} studied three-dimensional distributions of ACs and T2Cs in the context of the stellar evolution theory. They found that T2Cs are probably members of old and intermediate-age populations. The results for ACs are not straightforward, but the authors point out that these stars seem to belong to the old population as is demonstrated by their spread on-sky view.

\begin{table*}[htb]
\caption{Magellanic Bridge classical Cepheids: basic parameters}
\label{tab:ccs-gen}
\centering
\begin{tabular}{clDDDCCD@{ $\pm$}DD@{ $\pm$}D}
\multirow{2}{*}{Mode} & \multicolumn{17}{l}{OCVS ID} \\
{                   } & Loc. ID & \multicolumn2c{$P\ [{\rm d}]$} & \multicolumn2c{$I\ [{\rm mag}]$} & \multicolumn2c{$V\ [{\rm mag}]$} & {\rm RA} & {\rm Dec} & \multicolumn{4}{c}{$d\ [{\rm kpc}]^{({\it a})({\it b})}$} & \multicolumn{4}{c}{Age $[{\rm Myr}]^{({\it c})}$} \\

\hhline{=================}
\decimals
\multirow{10}{*}{F} & \multicolumn{17}{l}{OGLE-SMC-CEP-4956} \\
{                } &  M1 &  1.1162345 & 17.372 & 17.930 & 03^{\rm h}23^{\rm m}24\fs90 & -74\arcdeg58\arcmin07\farcs3 & 71.53 & 2.00 & 283 & 59 \\ \cdashline{2-17}
{                } & \multicolumn{17}{l}{OGLE-SMC-CEP-4953} \\
{                } &  M4 & 21.3856352 & 12.965 & 13.824 & 02^{\rm h}20^{\rm m}49\fs46 & -73\arcdeg05\arcmin08\farcs3 & 53.28 & 1.49 &  27 & 6  \\ \cdashline{2-17}
{                } & \multicolumn{17}{l}{OGLE-SMC-CEP-4952$^{(d)}$} \\
{                } &  M7 &  1.6414839 & 16.901 & 17.535 & 02^{\rm h}04^{\rm m}09\fs38 & -77\arcdeg04\arcmin38\farcs4 & 69.99 & 1.97 & 209 & 44 \\ \cdashline{2-17}
{                } & \multicolumn{17}{l}{OGLE-SMC-CEP-4987$^{(e)}$} \\
{                } & M10 &  2.9284749 & 15.738 & 16.458 & 03^{\rm h}31^{\rm m}34\fs40 & -70\arcdeg59\arcmin38\farcs2 & 56.45 & 1.56 & 132 & 28 \\ \cdashline{2-17}
{                } & \multicolumn{17}{l}{OGLE-SMC-CEP-4986$^{(f)}$} \\
{                } & M11 & 16.4454990 & 13.480 & 14.378 & 02^{\rm h}02^{\rm m}59\fs72 & -74\arcdeg03\arcmin24\farcs7 & 54.87 & 1.53 &  34 & 8 \\

\cline{1-17}
\multirow{8}{*}{1O} & \multicolumn{17}{l}{OGLE-SMC-CEP-4955} \\
{                 } &  M5 &  2.0308924 & 15.675 & 16.281 & 02^{\rm h}42^{\rm m}28\fs88 & -74\arcdeg43\arcmin17\farcs6 & 59.58 & 1.64 & 120 & 20 \\ \cdashline{2-17}
{                 } & \multicolumn{17}{l}{OGLE-LMC-CEP-3377} \\
{                 } &  M6 &  3.2144344 & 14.629 & 15.291 & 04^{\rm h}04^{\rm m}28\fs88 & -75\arcdeg04\arcmin47\farcs1 & 48.38 & 1.34 &  74 & 13 \\ \cdashline{2-17}
{                 } & \multicolumn{17}{l}{OGLE-LMC-CEP-3380} \\
{                 } & M12 &  1.0178714 & 16.485 & 17.101 & 04^{\rm h}35^{\rm m}32\fs89 & -74\arcdeg33\arcmin46\farcs7 & 53.62 & 1.48 & 252 & 41 \\ \cdashline{2-17}
{                 } & \multicolumn{17}{l}{OGLE-LMC-CEP-3381$^{(g)}$} \\
{                 } & M13 &  0.5188341 & 17.230 & 17.677 & 04^{\rm h}37^{\rm m}03\fs69 & -74\arcdeg58\arcmin25\farcs3 & 53.84 & 1.49 & 519 & 84 \\

\cline{1-17}
\multirow{2}{*}{1O2O} & \multicolumn{17}{l}{OGLE-SMC-CEP-4951$^{(g)}$} \\
{                   } & M9  &  0.7170500 & 16.769 & 17.222 & 02^{\rm h}02^{\rm m}33\fs88 & -75\arcdeg30\arcmin48\farcs0 & 54.06 & 1.49 & 367 & 60 \\

\cline{1-17}
\multicolumn{18}{p{.86\textwidth}}{All Cepheids except M1 and M7, form a continuous-like connection between the Magellanic Clouds. ({\it a}) The distance uncertainty does not include the mean LMC distance uncertainty from \citet{Pietrzynski2019} $d_{\rm LMC}=49.59\pm0.09\ {\rm(statistical)}\pm0.54\ {\rm(systematic)}\ {\rm kpc}$. ({\it b}) For comparison of distance estimates using different techniques see Tab.~\ref{tab:ccs-dist}. ({\it c}) This age determination was estimated using period-age relation from \citet{Bono2005}. For other estimates see Tab.~\ref{tab:ccs-ages}. ({\it d}) This Cepheid was reclassified from first-overtone to fundamental mode pulsator. ({\it e}) This Cepheid was added to the sample by \citet{Soszynski2017}. ({\it f}) $V$-band magnitude for this Cepheid was calculated using ASAS-SN Sky Patrol \citep{Schappee2014,Kochanek2017}. ({\it g}) Ages of short-period Cepheids may not be calculated properly (see details in Section~\ref{sec:ccs-ages}).}
\end{tabular}
\end{table*}

\begin{figure}[htb]
	\begin{center}
	\includegraphics[width=.47\textwidth]{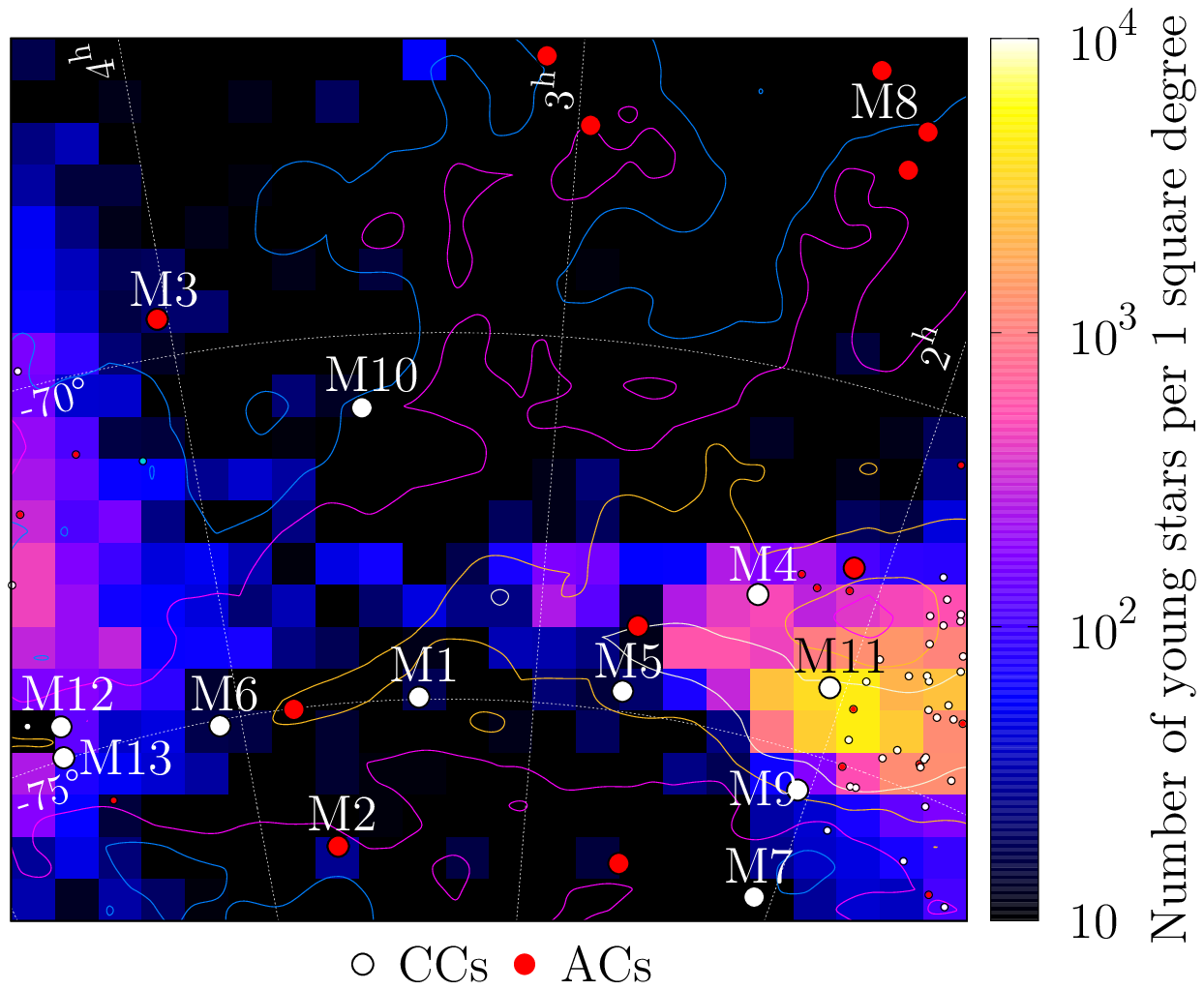}
	\end{center}
	\caption{On-sky locations of central Bridge Cepheid sample as compared to the color-coded young stars column density from \citet{Skowron2014} and neutral hydrogen density contours from Galactic All Sky \textsc{H i} Survey \citep{McClure-Griffiths2009,Kalberla2010,Kalberla2015}. Different types of Cepheids are marked with different colors. The selected Bridge sample is featured with larger dots, while smaller dots show LMC and SMC Cepheids. Labels M1-M9 mark classical Cepheid sample from \citealt{PaperI} and M10-M13 are new classical Cepheids that we added to the final MBR sample. M2, M3 and M8 were lately reclassified as anomalous Cepheids. The \textsc{H i} is integrated over velocity range $80<v<400\ {\rm km\ s^{-1}}$. Contours are on the levels $(1,2,4,8)\cdot10^{20}\ {\rm cm^{-2}}$. The color-coded value of each box is a logarithm of number of young stars per square degree area (each pixel is $\approx0.335$ square degrees). The map is represented in a Hammer equal-area projection centered at $\alpha_{\rm cen}=3^{\rm h}18^{\rm m}$, $\delta_{\rm cen}=-70\arcdeg$. This plot is an updated version of Fig.~18 from \citealt{PaperI}.}
	\label{fig:cep-onsky}
\end{figure}

In Fig.~\ref{fig:cep-onsky} we compare the on-sky distribution of different tracers in the central Bridge area. The plot shows classical (white dots), anomalous (red dots) and type II (green dots) Cepheids compared to the young stars distribution from \citet{Skowron2014}, which is color-coded, and neutral hydrogen density contours from Galactic All Sky \textsc{H i} Survey \citep{McClure-Griffiths2009,Kalberla2010,Kalberla2015}. Larger dots distinguish the selected Bridge sample, while smaller dots show other Magellanic System Cepheids. Note that there is only one T2C in the highlighted area. Labels M1--M13 mark classical Cepheids sample from \citealt{PaperI} as well as new classical Cepheids that we added to the final Bridge sample. Note that three of these objects were reclassified as ACs.

\subsection{Two- and Three-Dimensional Analysis}

The on-sky locations of CCs in the MBR are presented using white large dots in Fig.~\ref{fig:cep-onsky}. Their locations are matching very well the \textsc{H i} density contours. Only two Cepheids, namely M7 and M10, lie slightly offset from the peak \textsc{H i} density, though still well within contours showing the densest regions. Actually, the MBR CCs are forming an on-sky connection between the Magellanic Clouds following young stars' distribution \citep{Skowron2014}. Based on the on-sky locations, we conclude that all of our CCs in the Bridge match results from \citealt{PaperI} where we stated that the CCs add to the overall distribution of young population. For comparison we also show in Fig.~\ref{fig:cep-onsky} ACs which are marked with large red dots. ACs are definitely more spread and do not follow the young stars distribution, as was also already shown by other studies \citep{Fiorentino2012,Iwanek2018}.

\begin{figure*}[htb]
	\centering
	\includegraphics[width=.7\textwidth]{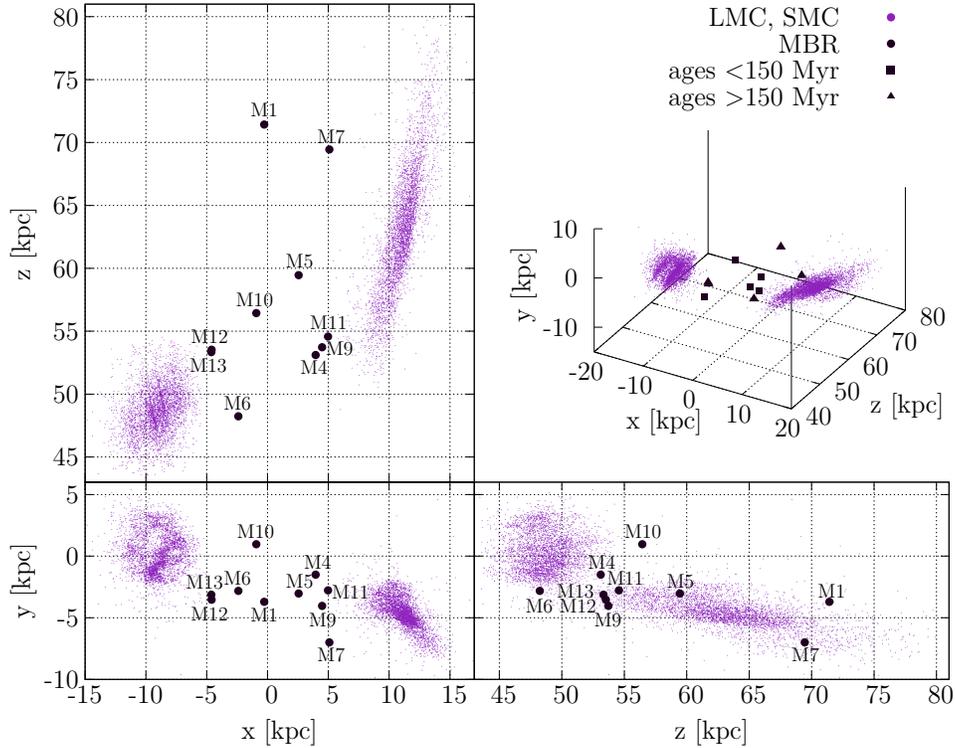}
	\caption{Three-dimensional distribution of classical Cepheids in the Magellanic System with MBR sample marked with large dots. Labels show local IDs of these objects (see Tab.~\ref{tab:ccs-gen}). The map is represented in the Cartesian coordinates with observer located at $(0,0,0)$. Ages were calculated using relations from \citet{Bono2005}.}
	\label{fig:ccs-3d}
\end{figure*}

Fig.~\ref{fig:ccs-3d} shows three-dimensional distribution of CCs in the Magellanic System. Four out of five CCs that we listed in \citealt{PaperI} as constituting a genuine connection between the Magellanic Clouds, specifically M4, M5, M6 and M9, have not been reclassified and their locations are the same as we presented therein. One out of these five, M3, was reclassified as AC. The four CCs that were lately added to the sample, M10-M13, add to the bridge-like structure. However, M12 and M13 may plausibly not belong to the genuine Bridge population as they seem to be the LMC outliers located in the extended LMC structure. Similarly, M9 and M11 are located very close to the SMC Wing and thus may be as well the Wing stars. On the other hand, the four LMC/SMC outliers may as well add to the main MBR sample. Taking that into account, we report here that eight out of ten CCs in our updated sample contribute to a bridge-like connection between the Magellanic Clouds.

The farthest CCs in our sample are M1 and M7. M7 is one of the two CCs that are located slightly offset from the \textsc{H i} contours and young population density distribution (see Fig.~\ref{fig:cep-onsky}). This suggests that M7 and M10 may have different origin than CCs discussed in previous paragraph. Yet, still they may constitute the genuine Bridge population. To test that, other parameters than discussed in this paper need to be taken into account (i.e. chemical composition). However, these Cepheids could also be members of the Counter Bridge, predicted by numerical model by \citet{Diaz2012}. This structure was already discussed in \citealt{PaperI} in terms of three-dimensional distribution of our previous sample, where we classified two CCs as plausible members of the Counter Bridge. Both were reclassified -- one (M8) as AC, another (M7) -- from fundamental mode CC to first-overtone pulsator (Sec.~\ref{sec:recl}). With the updated sample we do not have as evident candidates as before, though M1 and M7 are located near the borders of Counter Bridge (see Fig.~17 in \citealt{Ripepi2017}).

Our Bridge sample is not as spread in terms of distances as the sample presented in \citealt{PaperI}. All of the CCs are located in between the Magellanic Clouds, being farther than the closest LMC Cepheid and closer than the farthest SMC Cepheid. On the other hand, not all of the Bridge CCs form an evident, bridge-like connection. Some of these stars may as well be ejected from the LMC and/or SMC instead of forming the genuine Bridge. Indeed, we do see some individual objects spread over in different directions near these galaxies. The origin of our Bridge CCs will not be fully understood until further analysis are carried out taking into account different parameters than the ones we present in this paper. Of special importance are spectroscopic observations which could lead to a definite classification of these objects.

\subsection{Ages} \label{sec:ccs-ages}

Ages of our CCs were estimated using the period--age relation from \citet{Bono2005}. As we have already discussed in \citealt{PaperI} (see Section~6 therein), the Bridge has metallicity similar or smaller than SMC \citep{Lehner2008,Misawa2009}. \citet{Bono2005} do not provide any relation for metallicity smaller than SMC, thus we applied to our Bridge sample relation for the SMC metallicity. Calculated values are presented in Tabs.~\ref{tab:ccs-gen} and \ref{tab:ccs-ages}.

\begin{table*}[htb]
\caption{Magellanic Bridge classical Cepheids: ages}
\label{tab:ccs-ages}
\centering
\begin{tabular}{clDD@{ $\pm$}DD@{ $\pm$}DD}

Mode & Loc. ID & \multicolumn2c{$P\ [{\rm d}]^{(a)}$} & \multicolumn{4}{c}{${\rm Age}_{\rm PA} [{\rm Myr}]^{(b)}$} & \multicolumn{4}{c}{${\rm Age}_{\rm PAC} [{\rm Myr}]^{(c)}$} & \multicolumn2c{${\rm Age}_{\rm rot} [{\rm Myr}]^{(d)}$} \\
\hhline{==============}

\multirow{5}{*}{F} &           M1 &  1.1 & 283 & 59 & 271 & 63 & 567 \\ \cdashline{2-14}
{                } &           M4 & 21.4 &  27 & 6  &  27 & 8  &  48 \\ \cdashline{2-14}
{                } &           M7 &  1.6 & 209 & 44 & 207 & 50 & 410 \\ \cdashline{2-14}
{                } &          M10 &  2.9 & 132 & 28 & 110 & 26 & 252 \\ \cdashline{2-14}
{                } &          M11 & 16.4 &  34 & 8  &  35 & 10 &  59 \\
\cline{1-14}

\multirow{4}{*}{1O} &          M5 &  2.0 & 120 & 20 & 123 & 22 &  297 \\ \cdashline{2-14}
{                 } &          M6 &  3.2 &  74 & 13 &  79 & 15 &  191 \\ \cdashline{2-14}
{                 } &         M12 &  1.0 & 252 & 41 & 279 & 50 &  576 \\ \cdashline{2-14}
{                 } & M13$^{(e)}$ &  0.5 & 519 & 84 & 475 & 77 & 1101 \\
\cline{1-14}

               1O2O &  M9$^{(e)}$ &  0.7 & 367 & 60 & 329 & 54 &  807 \\
\cline{1-14}
\multicolumn{14}{p{.65\textwidth}}{({\it a}) Find a more precise period determination in Tab.~\ref{tab:ccs-gen}. ({\it b}) Calculated using period--age relation from \citet{Bono2005}. ({\it c}) Calculated using period--age--color relation from \citet{Bono2005}. ({\it d}) Calculated using period--age relation from \citet{Anderson2016} that includes stellar rotation. ({\it e}) Ages of short-period Cepheids may not be calculated properly (see details in Section~\ref{sec:ccs-ages}).}
\end{tabular}
\end{table*}

Eight out of ten CCs in our Bridge sample are younger than 300 Myr. This places them well within the context of MBR formation that occurred about 300 Myr ago, after the last encounter of the Magellanic Clouds (e.g. \citealt{Gardiner1994,Gardiner1996,Ruzicka2010,Diaz2012,Besla2012,Zivick2019}). Among these eight, six are constituting a connection between LMC and SMC, as we have described in the previous Section. These are CCs indexed M4, M5, M6, M10, M11, M12 and all except one are younger than 135 Myr. The oldest one in this subsample has the age of 252 Myr and is probably the LMC outlier.

Three CCs in our sample are younger than 75 Myr. These are M4 (27 Myr) which is located close to the SMC, M10 (34 Myr) and M6 (74 Myr) located in between the Magellanic Clouds. The young ages of all of the eight CCs younger than 300 Myr suggest that these were formed \textit{in-situ}. Especially the CCs located farther from the LMC and SMC and younger were plausibly formed in the genuine Bridge. However, we must also take into account the fact that these may be as well stars ejected from the LMC or SMC.

The two oldest CCs in our sample, M9 and M13, are also the shortest-period pulsators. M9 age determination is 367 Myr and M13 -- 519 Myr. While the former is reasonable for a Cepheid, the latter seems rather large. Actually both values could be incorrect due to the fact that models do not predict ages of objects with such short periods. That is why we treat these estimates as rather rough.

Tab.~\ref{tab:ccs-ages} presents the already discussed estimates based on period--age (PA) relation from \citet{Bono2005} as well as values obtained using their period--age--color (PAC) relation (we used relations for the SMC metallicity). In some cases the PA estimate is higher, in other the PAC. Even though, results from both relations match very well within the error bars. In the last column of Tab.~\ref{tab:ccs-ages} we also present age estimates using period--age relations from \citet{Anderson2016}. These relations were derived from models including rotation. Age values that they provide are approximately twice as large as values obtained using \citet{Bono2005} relations. This should not be surprising, as rotation induces mixing in stellar interiors which leads to refreshing the core hydrogen supplies. Thus, a rotating star can be burning hydrogen for longer time than a non-rotating one. As a result, the star can remain on the main sequence for a longer period of time, then cross the instability strip and become a Cepheid at an older age.

Even including rotation, five out of ten CCs in our Bridge sample have ages less than 300 Myr. This is still in agreement with an assumption that these objects were formed \textit{in-situ} after the last encounter of the Magellanic Clouds.

\subsection{Proper motions} \label{sec:ccs-pms}

\begin{figure}[htb]
	\centering
	\includegraphics[width=.47\textwidth]{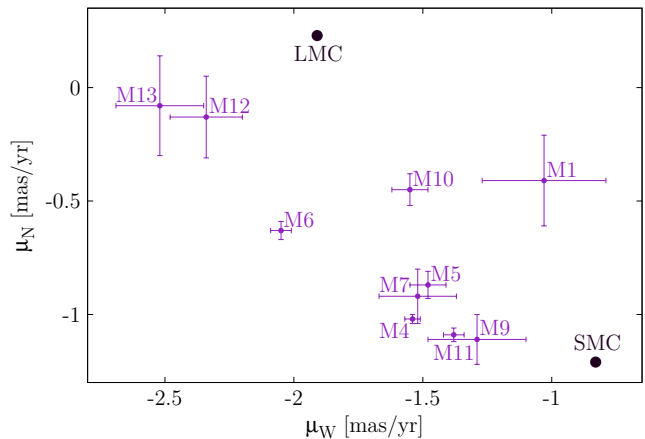}
	\caption{Proper motions of Bridge CCs as compared to the LMC \citep{Kallivayalil2013} and SMC \citep{Zivick2018}. All ten CCs from our sample are marked with their local IDs.}
	\label{fig:ccs-pms}
\end{figure}

\begin{figure}[htb]
	\centering
	\includegraphics[width=.47\textwidth]{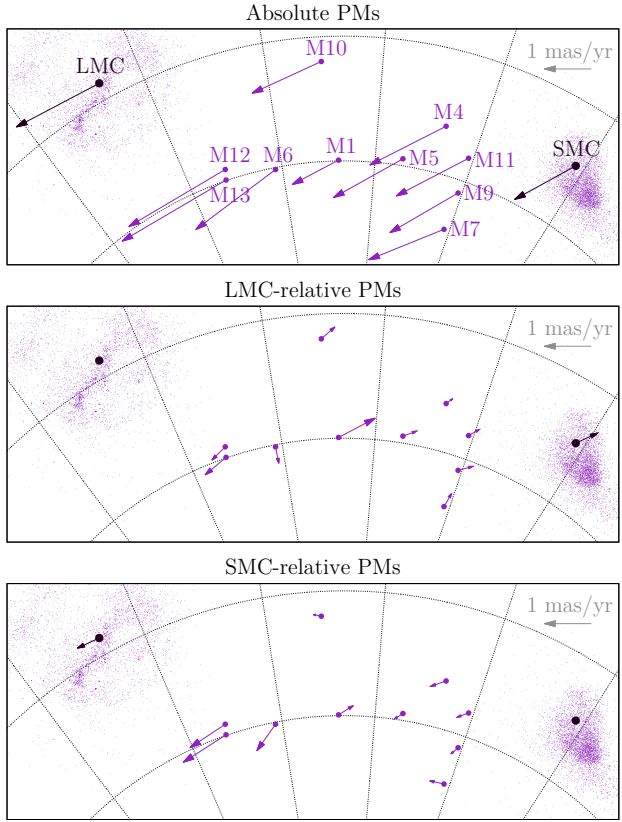}
	\caption{Proper motions of Bridge CCs as well as LMC \citep{Kallivayalil2013} and SMC \citep{Zivick2018} shown as vectors on the sky. Top panel presents absolute proper motions, while middle and bottom -- the LMC and SMC related frame, respectively. We adopted the LMC center of \citet{vanderMarel2014} and the SMC center of \citet{Stanimirovic2004}.}
	\label{fig:ccs-pmmap}
\end{figure}

We used {\it Gaia} DR2 \citep{Gaia2018} to analyse proper motions (PMs) of our Bridge CCs. Following \citet{Kallivayalil2013} and \citet{Zivick2018,Zivick2019} we use here $\mu_N=\mu_\delta$ and $\mu_W=-\mu_\alpha \cos \delta$, where $\alpha,\ \delta$ are RA, Dec, respectively. We compare our results to the LMC and SMC PMs \citep{Kallivayalil2013,Zivick2018} in Figs.~\ref{fig:ccs-pms} and \ref{fig:ccs-pmmap}. CCs PMs follow the general on-sky movement of the Magellanic System. PMs of M12 and M13 are relatively very similar to the LMC PM, while PMs of M9 and M11 -- to the SMC PM. This supports our conclusions from previous subsection that these Cepheids are probably LMC and SMC outliers. All of the other Bridge CCs PMs values fall in between those of LMC and SMC. This is what we would expect for a Bridge population (see Fig.~3 in \citealt{Zivick2019}).

Fig.~\ref{fig:ccs-pmmap} shows PMs of Bridge CCs as well as the LMC and SMC PMs plotted as vectors on the sky. CCs PMs as related to the LMC or SMC are rather low and comparable to the Clouds' relative PM. In the LMC-related frame all CCs except M12 and M13 are moving away from this galaxy. For the SMC-related PMs the situation is similar, as all CCs except M1 are also pointing away from this galaxy. This means that the Bridge CCs are moving away from both Clouds. M1 which is the only one moving towards the SMC, is also the farthest CC in our Bridge sample and is a Counter Bridge candidate.

\subsection{Different Distance Estimates} \label{sec:ccs-dist}

\begin{figure}[htb]
	\includegraphics[width=.47\textwidth]{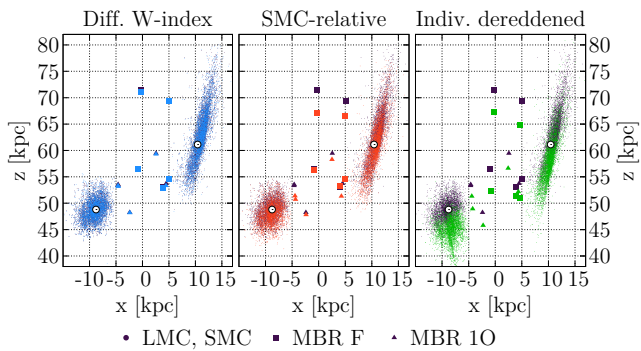}
	\caption{Three-dimensional distribution of CCs in the Magellanic System in Cartesian $xz$ plane projection. The distribution obtained using our basic distance estimates (as described in Section~\ref{sec:plr-dist}) is marked with black on every panel. Over plotted are, for comparison, different distributions marked with coloured dots (see text for details). The Bridge CCs sample is highlighted with larger marks. On all of the panels white circles mark LMC \citep{Pietrzynski2019,vanderMarel2014} and SMC \citep{Graczyk2014,Stanimirovic2004} centers.}
	\label{fig:ccs-dist}
\end{figure}

The Cepheid period--luminosity relation has an intrinsic dispersion caused by a finite width of the instability strip (e.g. \citealt{Anderson2016}) and/or depth effects (e.g. \citealt{Inno2013,Scowcroft2016,PaperI}). This implies that the PL relations are more useful for estimating sample's mean distance than individual distances of each Cepheid. The natural spread of PL relations is significantly smaller in infrared (e.g. \citealt{Storm2011,Ngeow2015,Scowcroft2016,Gallene2017,Madore2017}). However, one can obtain useful PL relations in optical regime with Wesenheit magnitude that combines two passbands and includes a color term \citep{Udalski1999,Fouque2007,Soszynski2008,Ngeow2012,Lemasle2013,Anderson2016,PaperI}. \citet{Ngeow2012} showed that the period--Wesenheit relations can be used to determine individual distances of Galactic Cepheids. Here we have also tried other techniques to calculate individual distances of our MBR CCs sample. The results are shown in Tab.~\ref{tab:ccs-dist} and Fig.~\ref{fig:ccs-dist} and discussed in this Section.

\begin{table*}[htb]
\caption{Magellanic Bridge classical Cepheids: distances}
\label{tab:ccs-dist}
\centering
\begin{tabular}{clDD@{ $\pm$}DD@{ $\pm$}DD@{ $\pm$}DD@{ $\pm$}D}

Mode & Loc. ID & \multicolumn2c{$P\ [{\rm d}]^{(a)}$} & \multicolumn{4}{c}{$d_{\rm LMC}\ [{\rm kpc}]^{(b)}$} & \multicolumn{4}{c}{$d_{\rm LMC, W44}\ [{\rm kpc}]^{(b)}$} & \multicolumn{4}{c}{$d_{\rm SMC}\ [{\rm kpc}]^{(c)}$} & \multicolumn{4}{c}{$d_{\rm red}\ [{\rm kpc}]$} \\
\hhline{====================}

\multirow{5}{*}{F} &  M1 &  1.1 & 71.53 & 2.00 & 71.17 & 1.89 & 67.22 & 1.86 & 67.37 & 10.83 \\ \cdashline{2-20}
{                } &  M4 & 21.4 & 53.28 & 1.49 & 53.00 & 1.41 & 53.43 & 1.50 & 51.53 &  7.46 \\ \cdashline{2-20}
{                } &  M7 &  1.6 & 69.99 & 1.97 & 69.87 & 1.87 & 66.98 & 1.85 & 65.40 & 10.25 \\ \cdashline{2-20}
{                } & M10 &  2.9 & 56.45 & 1.56 & 56.45 & 1.49 & 56.29 & 1.56 & 52.39 &  7.73 \\ \cdashline{2-20}
{                } & M11 & 16.4 & 54.87 & 1.53 & 54.80 & 1.46 & 55.30 & 1.55 & 51.24 &  7.36 \\
\cline{1-20}

\multirow{4}{*}{1O} &  M5 &  2.0 & 59.58 & 1.64 & 59.42 & 1.56 & 58.39 & 1.61 & 56.74 & 8.33 \\ \cdashline{2-20}
{                 } &  M6 &  3.2 & 48.38 & 1.34 & 48.31 & 1.27 & 47.95 & 1.33 & 45.91 & 6.40 \\ \cdashline{2-20}
{                 } & M12 &  1.0 & 53.62 & 1.48 & 53.61 & 1.40 & 51.66 & 1.43 & 49.18 & 7.51 \\ \cdashline{2-20}
{                 } & M13 &  0.5 & 53.84 & 1.49 & 53.47 & 1.40 & 51.01 & 1.41 & 51.62 & 8.21 \\
\cline{1-20}

               1O2O &  M9 &  0.7 & 54.06 & 1.49 & 53.66 & 1.41 & 51.63 & 1.42 & 52.50 & 8.14 \\
\cline{1-20}
\multicolumn{20}{p{.8\textwidth}}{({\it a}) Find a more precise period determination in Tab.~\ref{tab:ccs-gen}. ({\it b}) The distance uncertainty does not include the mean LMC distance uncertainty from \citet{Pietrzynski2019} $d_{\rm LMC}=49.59\pm0.09\ {\rm(statistical)}\pm0.54{\rm(systematic)}\ {\rm kpc}$. ({\it c}) The distance uncertainty does not include the mean LMC distance uncertainty from \citet{Graczyk2014} $d_{\rm SMC}=62.1\pm 1.9\ {\rm kpc}$.}
\end{tabular}
\end{table*}

As described in Section~\ref{sec:plr-dist}, our basic method of calculating distances is the same as we used in \citealt{PaperI}. It relies on Wesenheit PL relation for the LMC and an assumption that the fit corresponds to the mean LMC distance \citep{Pietrzynski2019}. We called this distance estimate $d_{\rm LMC}$, as it is related to the LMC, and show it in the third column in Tab.~\ref{tab:ccs-dist} (as well as in Tab.~\ref{tab:ccs-gen}). The resulting uncertainty does not include uncertainty from \citet{Pietrzynski2019}, as it would only lead to a systematic error, which would be the same for our entire sample. In order to test how the adopted reddening law influences individual distances, we also calculated distances the same way but with a different color term coefficient in the Wesenheit index. Instead of $1.55$ we used $1.44$ (see Equation~6 in \citealt{PaperI} and \citealt{Udalski2003}). The results are shown as $d_{\rm LMC, W44}$ (fourth column in Tab.~\ref{tab:ccs-dist}) and match very well our basic distances, although the former are slightly smaller. For comparison, see also left panel of Fig.~\ref{fig:ccs-dist} where the three--dimensional distribution obtained with basic distances is marked with black dots while with the different reddening law -- with blue -- and is overplotted on the former. This also means that the adopted reddening law does not have much impact on the Bridge Cepheids distances. This is in agreement with the fact that the reddening toward the Magellanic Bridge is low (\citealt{Schlegel1998,WagnerKaiser2017}, Skowron et al. in prep.).

We also calculated distances in relation to the SMC ($d_{\rm SMC}$, fifth column in Tab.~\ref{tab:ccs-dist}). We used the same technique as in our basic approach but adopted the SMC fit and the SMC mean distance as a reference \citep{Graczyk2014}. The resulting distances are smaller than our basic values and the difference is up to 5 kpc in some cases. Even though, the geometry of the entire LMC and SMC samples do not differ much using both approaches. This is shown in the middle panel of Fig.~\ref{fig:ccs-dist} where we overplotted the three-dimensional distribution relative to the SMC (red) on relative to the LMC (black). This incoherence may be caused by the fact that our SMC sample reveals a slightly larger mean distance when using our basic method than that from \citet{Graczyk2014}. Thus, when we changed the reference point to the SMC, the entire sample moved slightly closer.

Having magnitudes in both OGLE passbands, $I$ and $V$, we could also deredden our data. This is the same approach as used by \citet{Haschke2012a,Haschke2012b}. First, we calculated absolute magnitudes using PL relations from \citet{Sandage2004,Sandage2009} that were derived for the LMC and SMC data separately. We applied the SMC relations to the MBR sample, as the Bridge metallicity is close to or slightly lower than the SMC metallicity (e.g. \citealt{Lehner2008,Misawa2009,Carrera2017,WagnerKaiser2017}). We used relations not including the PL break at $\log P=1$, as the samples used to derive these relations only consisted of Cepheids with $\log P>0.4$. Half of our Bridge sample are CCs with shorter periods, thus we extrapolate these PL relations. Moreover, it was shown that the break at $\log P=1$ is not significant, at least for the SMC \citep{Bhardwaj2016}.

The PL relations that we used for the LMC \citep{Sandage2004}:
\begin{eqnarray}
M_I&=&(-2.949\pm0.020)\log P-(1.936\pm0.015) \\
M_V&=&(-2.701\pm0.035)\log P-(1.491\pm0.027)
\end{eqnarray}
And for the SMC \citep{Sandage2009}:
\begin{eqnarray}
M_I&=&(-2.862\pm0.028)\log P-(1.847\pm0.022) \\
M_V&=&(-2.588\pm0.045)\log P-(1.400\pm0.035)
\end{eqnarray}
These relations were derived only for the fundamental mode pulsators. For the first-overtone CCs in our sample we fundamentalized the periods using relation between periods from \citet{Alcock1995} (similarly as in \citealt{Groenewegen2000}):
\begin{equation}
P_{\rm 1O}/P_{\rm F} = 0.733-0.034\log P_{\rm F},\ 0.1<\log P_{\rm F}\leq0.7
\end{equation}
We have simplified the above equation and used in the following form:
\begin{equation}
P_{\rm F}=P_{\rm 1O}/(0.728-0.034\log P_{\rm 1O})
\end{equation}

It is noteworthy that relations for the LMC were derived using significantly different mean distance modulus to this galaxy. \citet{Sandage2004} based their calculations on value from \citet{Tammann2003}, which is $\mu_{\rm LMC}=18.54$ mag. In our basic approach we use $\mu_{\rm LMC}=18.477$ mag \citep{Pietrzynski2019}. For the SMC the difference is not that significant. \citet{Sandage2009} use $\mu_{\rm SMC}=18.93$ mag \citep{Tammann2008}, while \citet{Graczyk2014} obtain $\mu_{\rm SMC}=18.965$ mag.

Following \citet{Haschke2012a,Haschke2012b} approach, in the next step we calculated color excess for each Cepheid $E(V-I)=(m_V-m_I)-(M_V-M_I)$, where $m_{V,I}$ are observed magnitudes and $M_{V,I}$ -- absolute magnitudes in the appropriate filter. We noticed a mistake in \citet{Haschke2012a} Eqs.~6 and 7 that appears when trying to subtract one from another, and $A(V)-A(I)$ does not result in $E(V-I)$. We thus calculated these relations based on original \citet{Schlegel1998} coefficients to obtain total extinction in each passband:
\begin{eqnarray} \label{eqn:ext}
A_V&=&3.24(E(V-I)/1.278) \\
A_I&=&1.96(E(V-I)/1.278)
\end{eqnarray}
Note that there is 1.278 in the denominator instead of 1.4 as in \citet{Haschke2012a}. Calculated reddening parameters are shown in Tab.~\ref{tab:ccs-red} and discussed in the following Section, as here we concentrate on distances. To calculate distance moduli we used the $I$-band magnitudes as these values are usually more accurate than $V$-band. The distance modulus is simply:
\begin{equation}
\mu=m_I-M_I-A_I,
\end{equation}
and distance:
\begin{equation}
d=10^{(5+\mu)/5}.
\end{equation}

Results are presented in the last column of Tab.~\ref{tab:ccs-dist} and in the right panel of Fig.~\ref{fig:ccs-dist}. The individual dereddening technique resulted in significantly lower distances for every CC in the Bridge sample than previously discussed methods. Moreover, this technique has changed the entire geometry of the LMC and SMC samples, as is clearly visible in Fig.~\ref{fig:ccs-dist}. Our basic method relying on fitting the PL relations to the observational data is very robust, which was proven by many different surveys (e.g. \citealt{Haschke2012a,Haschke2012b, Moretti2014,PaperI,Inno2016,Ripepi2017}). Thus, we do not think that the individual dereddening technique is suitable to properly determine distances to Magellanic System Cepheids and especially infer any conclusions about structure and geometry.

\subsection{Reddening parameters}

\begin{table*}[htb]
\caption{Magellanic Bridge classical Cepheids: absolute magnitudes}
\label{tab:ccs-abs}
\centering
\begin{tabular}{lD@{ $\pm$}DD@{ $\pm$}DD@{ $\pm$}DD@{ $\pm$}DD@{ $\pm$}DD@{ $\pm$}D}
Loc. ID & \multicolumn{4}{c}{$M_I\ [{\rm mag}]^{(a)}$} & \multicolumn{4}{c}{$M_{I,2}\ [{\rm mag}]^{(b)}$} & \multicolumn{4}{c}{$M_V\ [{\rm mag}]^{(a)}$} & \multicolumn{4}{c}{$M_{V,2}\ [{\rm mag}]^{(b)}$} & \multicolumn{4}{c}{$E(V-I)\ [{\rm mag}]$} & \multicolumn{4}{c}{$E(V-I)_2\ [{\rm mag}]$} \\
\hhline{=========================}

  M1 & -1.984 & 0.028 & -1.742 & 0.184 & -1.524 & 0.036 & -1.175 & 0.209 & 0.098 & 0.053 & -0.009 & 0.280  \\
  M4 & -5.654 & 0.054 & -5.504 & 0.057 & -4.842 & 0.070 & -4.644 & 0.063 & 0.048 & 0.093 & -0.001 & 0.089  \\
  M7 & -2.463 & 0.029 & -2.233 & 0.159 & -1.957 & 0.037 & -1.628 & 0.181 & 0.128 & 0.054 &  0.029 & 0.242  \\
 M10 & -3.183 & 0.032 & -2.971 & 0.123 & -2.608 & 0.041 & -2.308 & 0.139 & 0.145 & 0.059 &  0.057 & 0.188  \\
 M11 & -5.327 & 0.051 & -5.170 & 0.052 & -4.547 & 0.065 & -4.336 & 0.058 & 0.118 & 0.087 &  0.072 & 0.082  \\
\cline{1-25}

  M5 & -3.140 & 0.032 & -2.928 & 0.125 & -2.569 & 0.041 & -2.268 & 0.141 & 0.035 & 0.059 & -0.053 & 0.191  \\
  M6 & -3.723 & 0.036 & -3.525 & 0.097 & -3.096 & 0.046 & -2.819 & 0.109 & 0.035 & 0.065 & -0.044 & 0.149  \\
 M12 & -2.264 & 0.028 & -2.029 & 0.170 & -1.777 & 0.036 & -1.440 & 0.192 & 0.129 & 0.054 &  0.027 & 0.258  \\
 M13 & -1.410 & 0.028 & -1.153 & 0.215 & -1.004 & 0.036 & -0.633 & 0.243 & 0.042 & 0.054 & -0.074 & 0.325  \\
\cline{1-25}

  M9 & -1.820 & 0.028 & -1.574 & 0.193 & -1.375 & 0.036 & -1.020 & 0.219 & 0.009 & 0.053 & -0.101 & 0.293  \\
\cline{1-25}
\multicolumn{25}{p{.97\textwidth}}{For first-overtone pulsators we used fundamentalized periods. ({\it a}) Calculated using relations from \citet{Sandage2004,Sandage2009}. ({\it b}) Calculated using relations from \citet{Gieren2018}.}
\end{tabular}
\end{table*}

Tab.~\ref{tab:ccs-abs} shows local IDs and absolute magnitudes in $I$ and $V$-bands, as well as color excesses of our Bridge CCs. For each passband we present two values for each parameter calculated using different PL relations \citep{Sandage2004,Sandage2009,Gieren2018}. As expected, the longer period the younger Cepheid, thus more luminous. Relations from \citet{Sandage2004,Sandage2009} have significantly different zero points than those of \citet{Gieren2018}, and it results in CCs being less luminous in the latter case. Relations from \citet{Gieren2018} also have larger uncertainties, and it is reflected in Tab.~\ref{tab:ccs-abs}. On the other hand, slopes are very consistent.

Color excesses, $E(V-I)$, in general have quite low values consistent with the fact that there is low extinction toward the Bridge area (\citealt{Schlegel1998,WagnerKaiser2017}, Skowron et al. in prep.). $E(V-I)$ calculated using relations from \citet{Gieren2018} in many cases have values that are physical only within the error bars, thus we use absolute magnitudes based on \citet{Sandage2004,Sandage2009} in further analysis. The discrepancy is probably due to a difference in zero points between these relations. However, we also note that relations from \citet{Gieren2018} were derived for CCs with periods $4<P<69$ d, and only three out of ten our CCs fall into this range.

Values obtained for color excesses of each CCs are very well consistent with the mean value of this parameter found toward the Bridge by \citet{WagnerKaiser2017}, who studied RRab type stars in that area. Their median is $E(V-I)=0.101\pm 0.007$ mag.

\begin{table*}[htb]
\caption{Magellanic Bridge classical Cepheids: reddening parameters}
\label{tab:ccs-red}
\centering
\begin{tabular}{lD@{ $\pm$}DD@{ $\pm$}DD@{ $\pm$}DD@{ $\pm$}DD@{ $\pm$}DD@{ $\pm$}D}
Loc. ID & \multicolumn{4}{c}{$A_I\ [{\rm mag}]^{(a)}$} & \multicolumn{4}{c}{$A_{I,W44}\ [{\rm mag}]^{(b)}$} & \multicolumn{4}{c}{$A_{I,{\rm t}}\ [{\rm mag}]^{(c)}$} & \multicolumn{4}{c}{$A_V\ [{\rm mag}]^{(a)}$} & \multicolumn{4}{c}{$A_{V,W44}\ [{\rm mag}]^{(b)}$} & \multicolumn{4}{c}{$A_{V,{\rm t}}\ [{\rm mag}]^{(c)}$} \\
\hhline{=========================}

  M1 &  0.083 & 0.070 &  0.094 & 0.067 & 0.248 & 0.134 & 0.181 & 0.073 & 0.192 & 0.071 & 0.150 & 0.081 \\
  M4 & -0.014 & 0.084 & -0.002 & 0.082 & 0.121 & 0.234 & 0.034 & 0.095 & 0.045 & 0.093 & 0.073 & 0.142 \\
  M7 &  0.139 & 0.070 &  0.143 & 0.070 & 0.325 & 0.137 & 0.267 & 0.074 & 0.271 & 0.072 & 0.196 & 0.083 \\
 M10 &  0.162 & 0.071 &  0.162 & 0.069 & 0.368 & 0.150 & 0.307 & 0.076 & 0.307 & 0.073 & 0.223 & 0.091 \\
 M11 &  0.111 & 0.082 &  0.113 & 0.080 & 0.319 & 0.221 & 0.236 & 0.091 & 0.239 & 0.090 & 0.193 & 0.134 \\
\cline{1-25}

  M5 & -0.061 & 0.071 & -0.055 & 0.068 & 0.089 & 0.149 & -0.025 & 0.075 & -0.019 & 0.073 & 0.054 & 0.090 \\
  M6 & -0.071 & 0.073 & -0.069 & 0.070 & 0.090 & 0.164 & -0.036 & 0.078 & -0.033 & 0.076 & 0.054 & 0.099 \\
 M12 &  0.102 & 0.069 &  0.103 & 0.067 & 0.327 & 0.135 &  0.231 & 0.073 &  0.232 & 0.070 & 0.198 & 0.082 \\
 M13 & -0.016 & 0.069 & -0.001 & 0.067 & 0.106 & 0.135 &  0.026 & 0.073 &  0.041 & 0.070 & 0.064 & 0.082 \\
\cline{1-25}

  M9 & -0.076 & 0.069 & -0.059 & 0.066 & 0.022 & 0.134 & -0.067 & 0.073 & -0.051 & 0.070 & 0.013 & 0.081 \\
\cline{1-25}
\multicolumn{25}{p{.97\textwidth}}{All parameters based on absolute magnitudes were calculated using relations from \citet{Sandage2004,Sandage2009} (see Tab.~\ref{tab:ccs-abs}). This is only an estimate and we discourage using values presented here in scientific research, as many obtained parameters are non-physical (values under zero). ({\it a}) Total reddening obtained using basic method distances. ({\it b}) Total reddening obtained using distances calculated assuming different reddening law (different color term coefficient in Wesenheit index as described in Section~\ref{sec:ccs-dist}). ({\it c}) Theoretical total reddening calculated without assuming any distance to each Cepheid. Here we used \citet{Schlegel1998} reddening laws (see Eq.~\ref{eqn:ext}).}
\end{tabular}
\end{table*}

Tab.~\ref{tab:ccs-red} presents reddening parameters for our Bridge CCs calculated using absolute magnitudes based on PL relations from \citet{Sandage2004,Sandage2009}. $A_{I,V}$ are total extinctions obtained using our basic method distances and $A_{(I,V),W44}$ are calculated using distances obtained with slightly different reddening law -- assuming different color term coefficient in the Wesenheit index (as described in Section~\ref{sec:ccs-dist}). Both values are very similar showing again that the adopted reddening law does not influence our technique much. However, the total extinction is of a quite low value, close to zero, and has a rather low precision (uncertainties are twice the obtained values or even higher). In some cases, the obtained value is even less than zero. We want to emphasise here that these values are not physical. We suggest that these values should not be used in any further study with the exception of making a comparison with a more accurate or different method.

Similarly to \citet{Haschke2012a,Haschke2012b}, we also calculated extinction without using \textit{a priori} distances but assuming a reddening law as described in Section~\ref{sec:ccs-dist} (see Eq.~\ref{eqn:ext}). Results are shown in Tab.~\ref{tab:ccs-red} as $A_{(I,V),t}$. Values obtained for $I$-passband are significantly larger than resulting from previously described methods, however, surprisingly, consistent within the error bars. The $V$-band extinction matches quite well with values obtained using other techniques. On the other hand, the error bars for $A_{(I,V),t}$ are quite high, thus we think that these results are not very useful.

The obtained results lead to a conclusion that calculating extinctions based on individual Cepheids may not be a relevant approach. Similarly, calculating distances based on these extinctions may as well lead to useless values.


\section{Anomalous Cepheids} \label{sec:acs}

\subsection{Final Sample and Basic Parameters}

We used the recently published OGLE Collection of ACs in the Magellanic System \citep{Soszynski2017} to construct our Bridge sample. Based on three-dimensional locations of these stars in comparison to the entire LMC and SMC samples, we decided to classify 10 ACs as Bridge candidates. Due to the latest updates and corrections applied to the OCVS (see Sec.~\ref{sec:ocvs}), three Bridge CCs were reclassified as ACs. That enlarged our ACs MBR sample to 13. Tab.~\ref{tab:acs} shows basic parameters of these objects: OCVS ID, local ID used in \citealt{PaperI} and this work (only for Cepheids reclassified from CCs to ACs), pulsation period $P$, magnitudes from both OGLE passbands ($I$ and $V$), Right Ascension and Declination (epoch J2000.0), distance $d$.

\begin{table*}[htb]
\caption{Magellanic Bridge anomalous Cepheids: basic parameters}
\label{tab:acs}
\centering
\begin{tabular}{clcDDDCCD@{ $\pm$}D}
Mode & OCVS ID & Loc. ID$^{({\it a})}$ & \multicolumn2c{$P\ [{\rm d}]$} & \multicolumn2c{$I\ [{\rm mag}]$} & \multicolumn2c{$V\ [{\rm mag}]$} & {\rm RA} & {\rm Dec} & \multicolumn{4}{c}{$d\ [{\rm kpc}]^{({\it b})}$} \\
\hhline{===============}

\multirow{11}{*}{F} & OGLE-LMC-ACEP-084                 & -- &  2.0506071 & 17.033 & 17.859 & 03^{\rm h}49^{\rm m}00\fs53 & -75\arcdeg00\arcmin49\farcs1 & 51.38 & 1.46 \\
{                 } & OGLE-LMC-ACEP-085                 & -- &  0.9156319 & 17.358 & 17.974 & 03^{\rm h}59^{\rm m}33\fs43 & -63\arcdeg16\arcmin40\farcs5 & 43.01 & 1.19 \\
{                 } & OGLE-SMC-ACEP-100                 & -- &  1.6414839 & 17.405 & 17.908 & 02^{\rm h}05^{\rm m}36\fs66 & -72\arcdeg24\arcmin19\farcs9 & 46.05 & 1.28 \\
{                 } & OGLE-SMC-ACEP-104                 & -- &  0.8780260 & 17.197 & 17.654 & 02^{\rm h}14^{\rm m}51\fs37 & -66\arcdeg59\arcmin30\farcs4 & 43.64 & 1.21 \\
{                 } & OGLE-SMC-ACEP-105                 & -- &  0.7559469 & 18.218 & 18.840 & 02^{\rm h}30^{\rm m}22\fs39 & -79\arcdeg08\arcmin25\farcs9 & 56.81 & 1.58 \\
{                 } & OGLE-SMC-ACEP-106                 & -- &  1.5007656 & 17.425 & 18.096 & 02^{\rm h}37^{\rm m}03\fs85 & -77\arcdeg03\arcmin02\farcs8 & 57.14 & 1.60 \\
{                 } & OGLE-SMC-ACEP-107                 & -- &  0.9317619 & 17.254 & 17.755 & 02^{\rm h}41^{\rm m}27\fs95 & -73\arcdeg48\arcmin45\farcs1 & 44.97 & 1.25 \\
{                 } & OGLE-SMC-ACEP-108                 & -- &  0.9147562 & 18.000 & 18.589 & 02^{\rm h}58^{\rm m}18\fs94 & -67\arcdeg05\arcmin46\farcs8 & 58.90 & 1.63 \\
{                 } & OGLE-SMC-ACEP-109                 & -- &  1.1701982 & 17.749 & 18.326 & 03^{\rm h}04^{\rm m}44\fs43 & -66\arcdeg11\arcmin15\farcs1 & 61.23 & 1.70 \\
{                 } & OGLE-LMC-ACEP-146$^{({\it c,d})}$ & M2 &  1.4300017 & 17.376 & 18.112 & 03^{\rm h}43^{\rm m}04\fs54 & -76\arcdeg56\arcmin02\farcs6 & 51.83 & 1.45 \\
{                 } & OGLE-GAL-ACEP-028$^{({\it c,e})}$ & M3 &  1.1589986 & 15.892 & 16.350 & 04^{\rm h}01^{\rm m}38\fs02 & -69\arcdeg28\arcmin40\farcs5 & 28.18 & 0.79 \\
\cline{1-15}

\multirow{3}{*}{1O} & OGLE-SMC-ACEP-102                 & -- &  0.9396136 & 17.347 & 17.904 & 02^{\rm h}13^{\rm m}39\fs52 & -66\arcdeg25\arcmin17\farcs0 & 58.35 & 1.67 \\
{                 } & OGLE-SMC-ACEP-120$^{({\it f})}$   & M8 &  0.8883309 & 17.302 & 17.738 & 02^{\rm h}21^{\rm m}28\fs45 & -65\arcdeg45\arcmin22\farcs4 & 60.05 & 1.72 \\
{                 } & OGLE-LMC-ACEP-147                 & -- &  0.7777591 & 16.537 & 16.961 & 04^{\rm h}35^{\rm m}35\fs29 & -81\arcdeg06\arcmin21\farcs0 & 39.01 & 1.13 \\
\cline{1-15}

\multicolumn{15}{p{\textwidth}}{({\it a}) Local IDs are provided only for ACs reclassified from CCs. ({\it b}) The distance uncertainty does not include the mean LMC distance uncertainty from \citet{Pietrzynski2019} $d_{\rm LMC}=49.59\pm0.09\ {\rm(statistical)}\pm0.54\ {\rm(systematic)}\ {\rm kpc}$. ({\it c}) These objects were reclassified from fundamental mode CCs. ({\it d}) Former OGLE-SMC-CEP-4957. ({\it e}) Former OGLE-LMC-CEP-3376. This Cepheid was reclassified as Milky Way object due to its proximity. ({\it f}) This object was reclassified from first-overtone CC. Former OGLE-SMC-CEP-4954.}
\end{tabular}
\end{table*}

To calculate individual distances of ACs we used the same technique as for classical Cepheids (Sec.~\ref{sec:plr-dist}). We applied one exception to $3\sigma$ clipping. We did not exclude one anomalous Cepheid from our sample that was treated by our algorithm as an outlier, namely OGLE-LMC-ACEP-147. This star is located in the newly added southern extension of the OGLE fields. The parameters of the fits are presented in Tab.~\ref{tab:acs-plr} and are consistent with those of \citet{Iwanek2018}. There is slight discrepancy between our results and those of \citet{Groenewegen2017} and \citet{Ripepi2014} that is probably caused by the latter being based on less numerous samples.

\begin{table*}[htb]
\caption{PL relations for ACs in the Magellanic System in the Wesenheit magnitude}
\label{tab:acs-plr}
\centering
\begin{tabular}{cccccccc}
\multicolumn{8}{c}{$W_{I,V-I}=a\log P+b$} \\
\cline{1-8}
Galaxy & P. mode & $a$ & $b$ [mag] & $\sigma$ [mag] & $\chi^2$/dof & $N_{\rm inc}$ & $N_{\rm rej}$ \\
\cline{1-8}

\decimals
\multirow{2}{*}{LMC} &  F & $-2.960\pm0.044$ & $16.599\pm0.007$ & 0.165 & 7.880 & 97 & 4 \\
{                  } & 1O & $-3.297\pm0.081$ & $16.041\pm0.017$ & 0.144 & 6.260 & 39 & 1 \\
\cline{1-8}
\multirow{2}{*}{SMC} &  F & $-2.725\pm0.054$ & $16.927\pm0.009$ & 0.178 & 9.228 & 74 & 1 \\
{                  } & 1O & $-3.710\pm0.094$ & $16.539\pm0.017$ & 0.169 & 8.592 & 40 & 0 \\
\cline{1-8}
\multicolumn{8}{p{.67\textwidth}}{F stands for fundamental mode, while 1O -- for first-overtone pulsators. $N_{\rm inc}$ is the number of objects included in the fit, while $N_{\rm rej}$ is the number of objects rejected during $3\sigma$-clipping procedure.}
\end{tabular}
\end{table*}

\subsection{Two- and Three-Dimensional Analysis}

The on-sky locations of all OGLE ACs along with CCs and T2Cs are presented in Fig.~\ref{fig:cep-all}, where the Bridge sample is highlighted with larger dots. Fig.~\ref{fig:cep-onsky} shows a close-up of the central Bridge area. The Cepheids locations are compared to young stars \citep{Skowron2014} and \textsc{H i} distribution (the Galactic All Sky \textsc{H i} Survey, \citealt{McClure-Griffiths2009,Kalberla2010,Kalberla2015}). Both plots clearly show that ACs are more spread than CCs and do not form as evident substructures as the latter in any area of the Magellanic System, including the Bridge. In contrary to CCs, ACs do not follow any line or bridge-like connection between the Clouds and do not match nor neutral hydrogen neither young population distribution. Nevertheless, this is what we could expect for an older stellar population. For a detailed statistical analysis of the three-dimensional distribution of ACs see \citet{Iwanek2018}.

\begin{figure}[htb]
	\includegraphics[width=.47\textwidth]{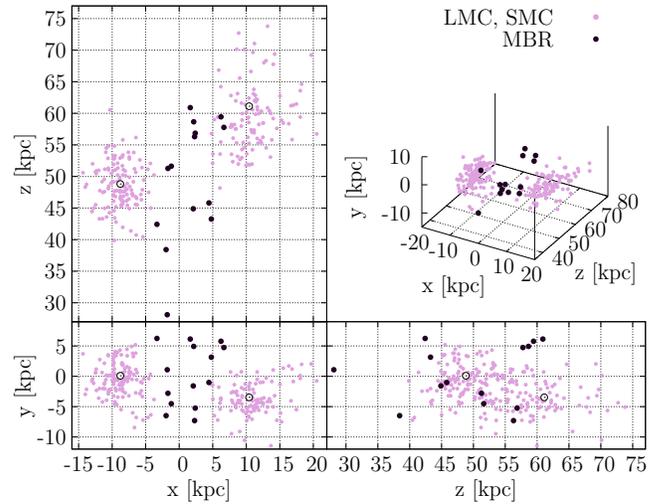}
	\caption{Three-dimensional distribution of anomalous Cepheids in the Magellanic System with MBR sample marked with darker dots. The map is represented in the Cartesian coordinates with observer located at $(0,0,0)$. White circles mark LMC \citep{Pietrzynski2019,vanderMarel2014} and SMC \citep{Graczyk2014,Stanimirovic2004} centers.}
	\label{fig:acs-3d}
\end{figure}

We were still able to distinguish the Bridge candidates located between the Magellanic Clouds in three-dimensions. Fig.~\ref{fig:acs-3d} shows three-dimensional distribution of ACs in the entire Magellanic System with the Bridge sample distinguished using larger dots. Although not very numerous, the ACs seem to create a rather smooth connection between the Clouds. However, we cannot state that this connection is bridge-like because these ACs may be as well LMC and/or SMC outliers that we also see located in different directions around these galaxies.

\subsection{Proper motions} \label{sec:acs-pms}

\begin{figure}[htb]
	\centering
	\includegraphics[width=.47\textwidth]{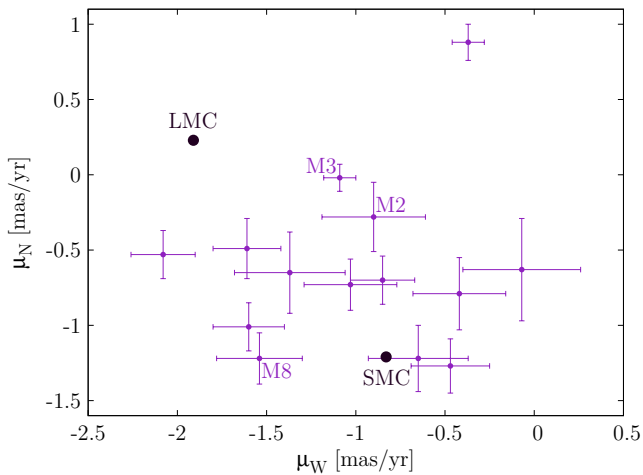}
	\caption{Proper motions of Bridge ACs as compared to the LMC \citep{Kallivayalil2013} and SMC \citep{Zivick2018}. Three reclassified Cepheids are shown with their local IDs.}
	\label{fig:acs-pms}
\end{figure}

\begin{figure}[htb]
	\centering
	\includegraphics[width=.47\textwidth]{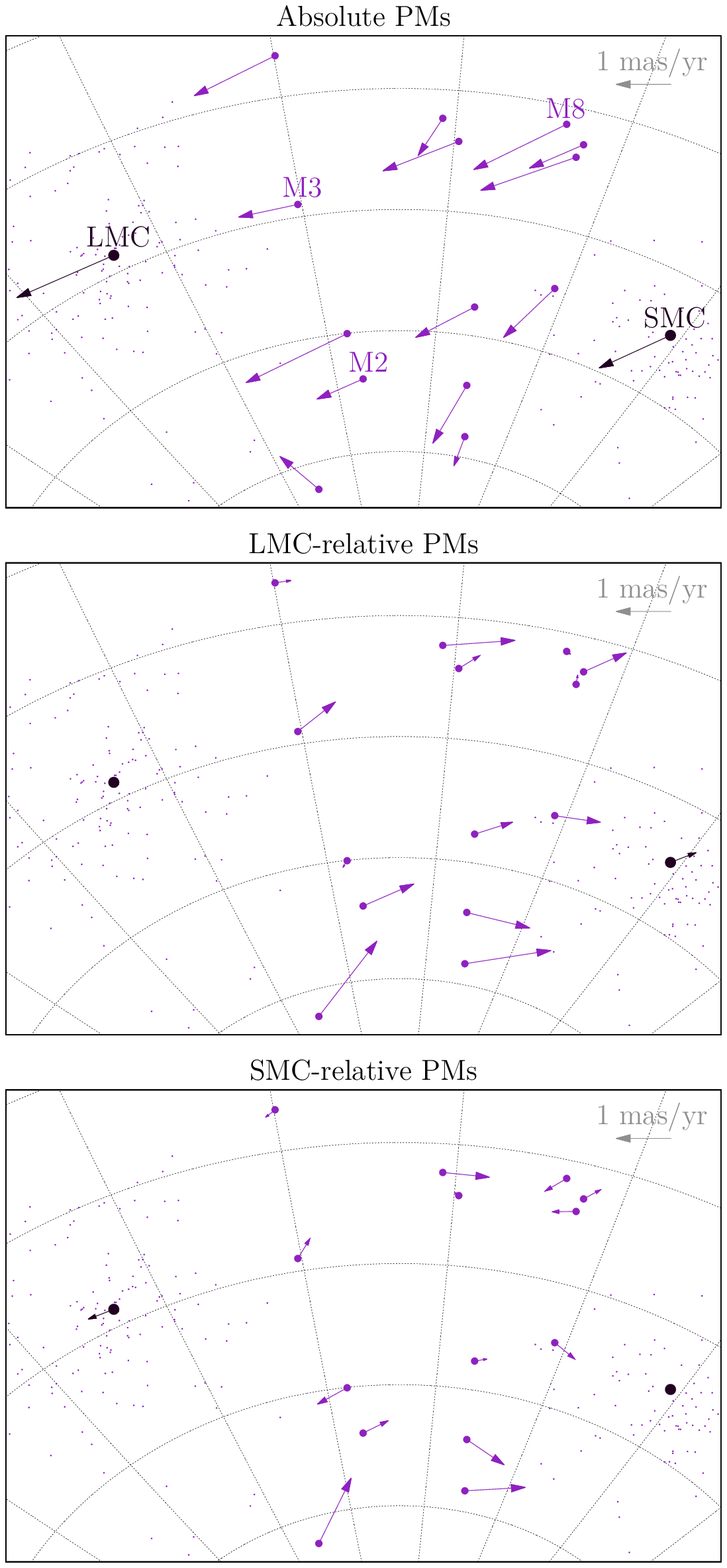}
	\caption{Proper motions of Bridge ACs as well as LMC \citep{Kallivayalil2013} and SMC \citep{Zivick2018} shown as vectors on the sky. Top panel presents absolute proper motions, while middle and bottom -- the LMC and SMC related frame, respectively. We adopted the LMC center of \citet{vanderMarel2014} and the SMC center of \citet{Stanimirovic2004}.}
	\label{fig:acs-pmmap}
\end{figure}

Similarly to CCs, we also used {\it Gaia} DR2 \citep{Gaia2018} to analyse PMs of our Bridge ACs. Again, we compare results to the LMC and SMC PMs in Figs.~\ref{fig:acs-pms} and \ref{fig:acs-pmmap}. ACs follow the general on-sky movement of the entire Magellanic System. Almost all of them fall into the PMs range that we would expect for a Bridge objects (see Fig.~3 from \citealt{Zivick2019}). The only exception is an object with $\mu_N$ close to 1 is located near to the southern edge of the OGLE fields in the Bridge area.

In the LMC relative frame of motion virtually all of the ACs move away from this galaxy. The SMC relative motions reveal that indeed more than half of our Bridge sample is approaching the SMC. The relative PMs are rather low and range from almost zero up to four times the relative motion of the Magellanic Clouds. Comparing ACs PMs to those of CCs the former reveal a more complicated and chaotic movements on the sky.

\section{Reclassified Cepheids} \label{sec:recl}

Latest reclassification of four CCs is slightly disputable, as all of these objects have parameters located close to the CCs/ACs (or CC F/1O) boundary. In Tab.~\ref{tab:cep-recl} we compare basic parameters of the four stars before and after the reclassification and list: local ID, type and mode as well as distance and age before and after the reclassification. The estimates for the latter were already presented in the previous Sections. The estimates for before the reclassification were calculated simply including these objects in the appropriate CCs or ACs sample and using the same technique as for the entire samples that we present in this paper.

\begin{table*}[htb]
\caption{Magellanic Bridge Cepheids: reclassification}
\label{tab:cep-recl}
\centering
\begin{tabular}{cl@{ $\to$ }lD@{ $\pm$}D@{ $\to$}D@{ $\pm$}DD@{ $\pm$}D@{ $\to$ }D@{ $\pm$}D}

\multirow{2}{*}{Loc. ID} & \multicolumn{18}{c}{Before $\to$ after} \\
\cdashline{2-19}
 & \multicolumn{2}{c}{Type and mode} & \multicolumn{8}{c}{$d\ [{\rm kpc}]^{({\it a})}$} & \multicolumn{8}{c}{Age $[{\rm Myr}]^{({\it b})}$} \\
\cline{1-19}
M2 &   CC F & AC F  & 74.07 & 2.08 & 51.83 & 1.45 & 233 & 49 & \multicolumn{4}{c}{NA} \\
M3 &   CC F & AC F  & 39.81 & 1.11 & 28.18 & 0.78 & 275 & 57 & \multicolumn{4}{c}{NA} \\
M8 &  CC 1O & AC 1O & 80.95 & 2.23 & 60.05 & 1.72 & 292 & 48 & \multicolumn{4}{c}{NA} \\
M7 &  CC 1O & CC F  & 88.83 & 2.45 & 69.99 & 1.97 & 151 & 25 & 209 & 44 \\
\cline{1-19}
\multicolumn{19}{p{.7\textwidth}}{({\it a}) The distance uncertainty does not include the mean LMC distance uncertainty from \citet{Pietrzynski2019} $d_{\rm LMC}=49.59\pm0.09\ {\rm(statistical)}\pm0.54\ {\rm(systematic)}\ {\rm kpc}$. ({\it b}) This age determination was estimated using period-age relation from \citet{Bono2005} and is available for CCs only.}
\end{tabular}
\end{table*}

Bottom row of Fig.~\ref{fig:cep-plr} shows the four reclassified Cepheids on the PL relations for the entire LMC (left panel) and SMC (right panel) CCs and ACs samples. The Bridge Cepheid sample is overplotted on each panel using large marks. Additionally, the reclassified Cepheids are also marked with a star and their local ID. We discuss locations of these objects on the PL diagrams according to all of the presented relations as these Cepheids may be neither LMC nor SMC members. Thus, their parameters need to be analyzed in a broader context. Note that we do not classify objects based only on their location on the PL diagrams but we mainly use their light curve (shape and Fourier decomposition parameters, \citealt{Soszynski2015a}).

M7, which was reclassified from first-overtone CC to fundamental mode CC is indeed located much closer to the fundamental mode than first-overtone PL relations. This object is also situated close to the LMC fundamental-mode ACs but at the same time close to the SMC fundamental mode CCs. M2, recently reclassified from fundamental mode CC to fundamental mode AC, is very close to the LMC fundamental mode ACs PL relation. On the other hand, it is located in between the fundamental mode CCs and ACs PL relations for the SMC. M3 is another object reclassified in the same way as M2. M3 is situated almost on the fit that we obtained for the first-overtone CCs in the LMC. In fact, it is located quite far from the LMC fundamental mode PL relation for the CCs and for the ACs -- even farther. Compared to the SMC relations M3 definitely seems to be an outlier from the fundamental mode PL relations. In the case of M8, which was reclassified from the first-overtone CC to the first-overtone AC the closest PL relations in the LMC are relations for both types of ACs. This star is located between these relations. When compared to the SMC M8 is situated close to the first-overtone PL relation for ACs but at the same time quite close to both PL relations for the CCs.

\begin{figure}[htb]
	\includegraphics[width=.47\textwidth]{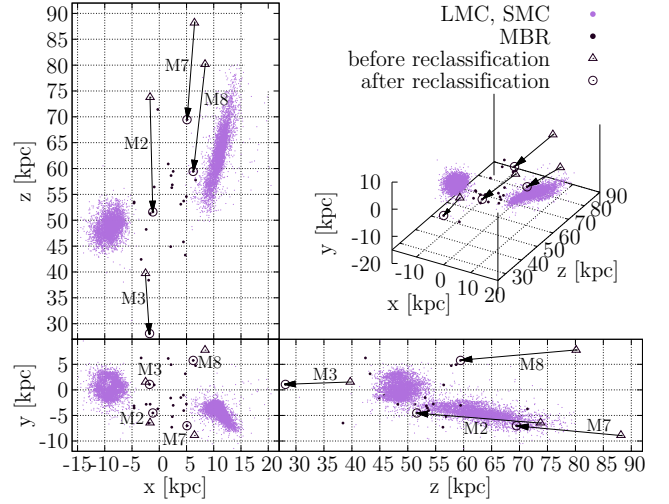}
	\caption{Three-dimensional distribution of CCs and ACs in the Magellanic System with MBR sample marked with large dots. Additionally, locations of four reclassified Cepheids are highlighted with different markers. Arrows show the direction of changes in locations. Labels show local IDs of these objects (see Tab.~\ref{tab:cep-recl}). The map is represented in the Cartesian coordinates with observer located at $(0,0,0)$.}
	\label{fig:cep-recl}
\end{figure}

The reclassification has significantly changed three-dimensional distribution of Cepheids in the Bridge area as distances of all reclassified objects have decreased by more than 10 kpc in each case. We show this change in Fig.~\ref{fig:cep-recl} where we plotted projections of three-dimensional Cartesian distribution of all Cepheids analyzed here (both CCs and ACs) with the Bridge sample highlighted using larger dots. The reclassified objects are marked separately and the arrows show the change of distances that occurred with the reclassification.

A change of close to or more than 20 kpc has occurred for M2, M7 and M8. If these stars were not reclassified they would be perfect candidates for Counter Bridge members, as we have already stated in \citealt{PaperI}. Moreover, their ages would match very well the scenario in which they would be formed \textit{in-situ} in this structure. M2 and M8 were reclassified as ACs and after this change these objects are located in between the Magellanic Clouds matching very well three-dimensional distribution of ACs (see Fig.~\ref{fig:acs-3d}). M7 is a CC and even after the reclassification this star could be a Counter Bridge member though it is now located farther from the center of this structure, and thus this scenario is less plausible (we have discussed M7 location in details in Section~\ref{sec:ccs-dist}).

In our Bridge CC sample from \citealt{PaperI} M3 was the closest Cepheid -- located even closer than any LMC CC. After the reclassification, this object is located even closer at $\sim28$ kpc -- halfway between the Sun and the Magellanic System. Due to this M3 was treated as LMC outlier by our $3\sigma$-clipping algorithm that we applied to the ACs sample. Based on its proximity, we decided to classify this object as Milky Was halo AC.


\section{{\it Gaia} DR2 Cepheids in the Bridge}

\subsection{Comparison with OCVS}

The {\it Gaia} DR2 contains a list of variable stars including Cepheids and RR Lyrae stars \citep{Gaia2018,Holl2018,Clementini2019}. As following \citet{Holl2018}, due to the probabilistic and automated nature of the classification process, the {\it Gaia} DR2 catalogue of classical variables is not as complete and pure as the OGLE Collection of Variable Stars is (see Tab.~2 in \citealt{Holl2018} and \citealt{Clementini2019}). In this Section we revive the {\it Gaia} DR2 classical pulsators, listed in \texttt{vari\_cepheid} table \citep{Gaia2018,Holl2018}, in the Magellanic Bridge area and compare it to the OCVS.

\begin{figure}[htb]
	\centering
	\includegraphics[width=.47\textwidth]{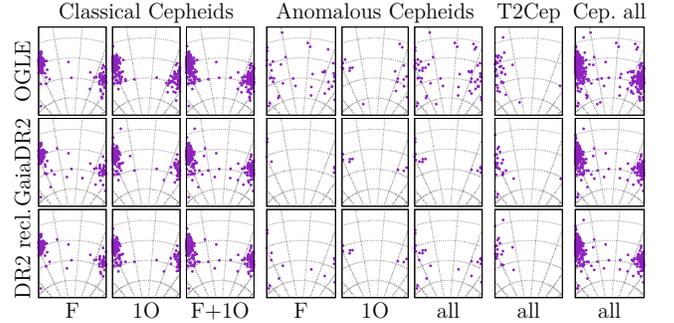}
	\flushleft
	\caption{A comparison of OGLE (top row) and {\it Gaia} DR2 (middle and bottom row) Cepheids in the Magellanic Bridge area. The bottom row shows DR2 sample after reclassification made by \citet{Ripepi2018}. It may seem that {\it Gaia} DR2 discovered more CCs in the Bridge area than contained in the nearly complete OCVS. However, a comparison of different panels leads to a conclusion that many of ACs were classified in DR2 as CCs. Finally, the OCVS contains several more Cepheids in the Bridge area than DR2.}
	\label{fig:ccs-dr2}
\end{figure}

Fig.~\ref{fig:ccs-dr2} compares on-sky locations of individual Cepheids of different types and modes in the Bridge area. The top row shows OGLE data, while the middle and bottom - {\it Gaia} DR2. The latter shows DR2 Cepheid sample after reclassification made by \citet{Ripepi2018}. First three columns show classical Cepheids of the following modes, both single and multi-mode -- fundamental, first-overtone and both of these together. Based on only these plots, it may seem that {\it Gaia} DR2 discovered several new classical Cepheids that were not present in the virtually complete OGLE Collection of Classical Cepheids \citep{Soszynski2017}.

Comparing distributions of anomalous Cepheids, both fundamental mode and first-overtone pulsators as well as entire samples, the {\it Gaia} DR2 seems to classify no objects as anomalous Cepheids in the Bridge. At the same time, the OCVS contains many ACs in between the Magellanic Clouds. This leads to a conclusion that many ACs were classified as CCs in DR2. This is probably due to different classification methods used in both cases (i.e. see reclassification of the Milky Way Cepheids from {\it Gaia} DR2 in \citealt{Ripepi2018}). It is very similar for T2Cs, though neither OGLE nor {\it Gaia} DR2 classify any objects of this type in the central Bridge area. A comparison of all of the Cepheids between the Magellanic Clouds reveals that the {\it Gaia} DR2 has incorrectly catalogued a number of objects in the Bridge area.

We compared the OCVS and {\it Gaia} DR2 Cepheid samples in numbers. For the cross-match we selected a DR2 sample covering the entire OGLE fields in the Magellanic System (see Fig.~\ref{fig:cep-all}). We use the OCVS sample containing the latest updates and corrections as described in Sec.~\ref{sec:ocvs}. Out of 10140 Cepheids included in the OGLE Collection in the Magellanic System (9532 CCs, 268 ACs, 340 T2Cs) 7490 objects were found in the {\it Gaia} DR2 Cepheid sample. Thus, when comparing to the virtually complete OGLE Collection of Cepheids, the {\it Gaia} DR2 completeness is on a level of 73.9\%, which is consistent with Tab.~2 in \citet{Holl2018}. High completeness is not surprising as the OCVS Cepheid data set from the Magellanic Clouds was a training set for the {\it Gaia} Cepheid detection algorithms. In other areas of the sky, the {\it Gaia} DR2 Cepheid sample completeness is significantly lower, i.e., \citet{Udalski2018} showed that in the Milky Way disk and bulge area it is on a level of 9.1\%.

We additionally compared the {\it Gaia} DR2 detections in the region designed as MBR in OGLE-IV fields (Fig.~18 in \citealt{Udalski2015}). 30 {\it Gaia} DR2 Cepheids are located in the OGLE MBR field footprint. 29 were confirmed in the OGLE Collection as genuine Cepheids and the one lacking object is likely an eclipsing star. 59 Cepheids in the OGLE Collection (CCs, ACs and T2Cs) lie in the OGLE MBR fields. Thus, the completeness of the {\it Gaia} DR2 in this region is $29 / 59 \simeq 49\%$.


\section{Conclusions}

In this paper, which is the third in a series of analyzing three-dimensional structure of the Magellanic System, we present an updated detailed analysis of Cepheids in the Magellanic Bridge. We use data from the OGLE project -- released parts of the OCVS \citep{Soszynski2015b,Soszynski2017,Soszynski2018} as well as data that were not yet published. The Collection was recently updated: seven Cepheids were added and four were reclassified. We present a thorough study of classical and anomalous Cepheid Bridge samples using very precise OGLE photometry and astrometry. We note that we did not classify any T2C as MBR member due to their absence in this area.

Similarly to \citealt{PaperI}, our basic method of calculating distances relies on fitting PL relations using Wesenheit $W_{I,V-I}$ index to the entire LMC sample. Then we estimate individual distance of each Cepheid relative to the LMC mean distance and the LMC fit. Based on three-dimensional coordinates as well as on-sky locations of stars in relation to the LMC and SMC entire samples, we selected our Bridge samples.

The updated Bridge CC sample contains 10 objects. As compared to \citealt{PaperI} sample, we removed three objects (M2, M3 and M8 that were reclassified as ACs) and added four objects (M10 added by \citealt{Soszynski2017}, M11-M13). On-sky locations of CC MBR sample match very well \textsc{H i} density contours and young stars distribution. Only two Cepheids, namely M7 and M10, are located slightly offset, though still well within the densest regions. The CCs add to the overall distribution of young stars in the Bridge area.

In three dimensions, eight out of ten objects from the CC sample form a bridge-like connection between the Magellanic Clouds. Four out of these eight are located close to the LMC (M12 and M13) or SMC (M9 and M11). Two that do not form the bridge-like connection, namely M1 and M7, are located slightly farther than the main sample, thus they may constitute a Counter Bridge. However, they may also be genuine MBR members. Further study is needed to test this. We also analyzed different methods of obtaining distances and conclude that the adopted reddening law does not influence results much and the reddening toward the Bridge is low. Moreover, the individual dereddening method used by i.e. \citet{Haschke2012a,Haschke2012b} seems to be inappropriate in this case.

Eight out of ten Bridge CCs have ages of less than 300 Myr (as based on the period--age relation from \citealt{Bono2005}). This agrees with a hypothesis that they were formed \textit{in-situ}. The three youngest CCs have ages less than 75 Myr. The two oldest CCs can be LMC or SMC members. Moreover, their periods are shorter than 1 day, thus their age estimate may not be appropriate as the models do not predict ages of such short period pulsators. We also tested period--age--color relations from \citet{Bono2005}. Obtained results match very well previous estimates. Using period--age relations including rotation \citep{Anderson2016} leads to twice as large values as without rotation. Even though, still five out of ten CCs in our Bridge sample have ages smaller than 300 Myr. This is in agreement with the statement that these objects were formed in the Bridge after the last encounter of the Magellanic Clouds.

Our Bridge AC sample consists of 13 objects. Their on-sky locations are not matching \textsc{H i} either young stars density contours. ACs distribution is very spread in both two and three dimensions. However, they form a rather smooth connection between the Magellanic Clouds. But also, we cannot state that this connection is bridge-like, as these stars may as well be LMC/SMC outliers.

We also tested {\it Gaia} DR2 Cepheids on-sky distribution in the Bridge area. DR2 contains more CCs in the MBR than the OCVS. However, DR2 does not include virtually any AC in between the Magellanic Clouds. This is explained by a different classification process, where many ACs are classified as CCs in DR2. A comparison of all types of Cepheids shows that the OCVS has more objects in the MBR, thus is definitely more complete.

We present a complementing study of older classical pulsators in the Magellanic Bridge -- RR Lyrae stars -- in a closely following \citealt{PaperIV}.


\acknowledgments

A.M.J.-D. is supported by the Polish Ministry of Science and Higher Education under ``Diamond Grant'' No. DI2013 014843 and by Sonderforschungsbereich SFB 881 "The Milky Way System" (subproject A3) of the German Research Foundation (DFG). The OGLE project has received funding from the National Science Centre, Poland, grant MAESTRO 2014/14/A/ST9/00121 to A.U.

We would like to thank all of those, whose remarks and comments inspired us and helped to make this work more valuable. In particular we would like to thank Richard Anderson, Abhijit Saha, Vasily Belokurov, Anthony Brown, Laurent Eyer, Martin Groenewegen, Vincenzo Ripepi, Rados\l{}aw Smolec, Martino Romaniello, Krzysztof Stanek.

This research was supported by the Munich Institute for Astro- and Particle Physics (MIAPP) of the DFG cluster of excellence "Origin and Structure of the Universe", as it benefited from the MIAPP program "The Extragalactic Distance Scale in the {\it Gaia} Era" as well as International Max Planck Research School (IMPRS) Summer School on "{\it Gaia} Data and Science 2018".

This work has made use of data from the European Space Agency (ESA) mission {\it Gaia} (\url{https://www.cosmos.esa.int/gaia}), processed by the {\it Gaia} Data Processing and Analysis Consortium (DPAC, \url{https://www.cosmos.esa.int/web/gaia/dpac/consortium}). Funding for the DPAC has been provided by national institutions, in particular the institutions participating in the {\it Gaia} Multilateral Agreement.


\end{document}